\documentclass{aa}  

\usepackage{graphicx}
\usepackage{txfonts}
\usepackage{subfigure}
\usepackage{xcolor} % Asegúrate de que el paquete esté incluido

\begin{document}

\titlerunning{LAE multiples}
%\authorrunning{name(s) of author(s)}
   
%\title{Looking for the impact of gravitational perturbations in Lyman-Alpha Emitter multiples}
\title{ODIN: Using multiplicity of Lyman-Alpha Emitters to assess star formation activity in dark matter halos}

%\subtitle{I. Overviewing the $\kappa$-mechanism}

\author{M. Candela Cerdosino\inst{1,2,3}\fnmsep\thanks{e-mail: candelacerdosino@mi.unc.edu.ar}
\and
Nelson Padilla\inst{1,3}
\and
Ana Laura O'Mill\inst{1,3}
\and
Eric Gawiser\inst{4}
\and
Nicole M. Firestone\inst{4} 
\and
Maria Celeste Artale\inst{5} 
\and
Kyoung-Soo Lee\inst{6}
\and
Changbom Park\inst{7}
\and
Yujin Yang\inst{8}
\and
Caryl Gronwall\inst{9,10} 
\and
Lucia Guaita\inst{5,11} %%
\and
Sungryong Hong\inst{8} 
\and
Ho Seong Hwang\inst{12,13,14} 
\and
Woong-Seob Jeong\inst{8}
\and
Ankit Kumar\inst{5}
\and
Jaehyun Lee\inst{8}
\and
Seong-Kook Joshua Lee\inst{13} 
\and
Paulina Troncoso Iribarren\inst{15} 
\and
Ann Zabludoff\inst{16} 
}

\institute{Instituto de Astronomía Teórica y Experimental (IATE), CONICET-UNC, Laprida 854, X500BGR, Córdoba, Argentina\
%\email{wuchterl@amok.ast.univie.ac.at} 1
\and
Facultad de Matemática, Astronomía, Física y Computación, Universidad Nacional de Córdoba, Bvd. Medina Allende s/n, Ciudad Universitaria, X5000HU, Córdoba, Argentina\ %2
\and
Observatorio Astronómico de Córdoba, Universidad Nacional de Córdoba, Laprida 854, X5000BGR, Córdoba, Argentina\ %3
\and
Department of Physics and Astronomy, Rutgers, the State University of New Jersey, Piscataway, NJ 08854, USA\ %4
\and
Universidad Andres Bello, Facultad de Ciencias Exactas, Departamento de Fisica y Astronomia, Instituto de Astrofisica, Fernandez Concha 700, Las Condes, Santiago RM, Chile\ %5
\and
Department of Physics and Astronomy, Purdue University, 525 Northwestern Avenue, West Lafayette, IN 47906, USA\ %6
\and
Korea Institute for Advanced Study, 85 Hoegi-ro, Dongdaemun-gu, Seoul 02455, Republic of Korea %7
\and
Korea Astronomy and Space Science Institute, 776 Daedeokdae-ro, Yuseong-gu, Daejeon 34055, Republic of Korea %8
\and 
Department of Astronomy and Astrophysics, The Pennsylvania State University University Park, PA 16803, USA\  %9
\and
Institute for Gravitation and the Cosmos, The Pennsylvania State University, University Park, PA 16803, USA\ %10
\and
Millennium Nucleus for Galaxies (MINGAL)\ %11
\and
Astronomy Program, Department of Physics and Astronomy, Seoul National University, 1 Gwanak-ro, Gwanak-gu, Seoul 08826, Republic of Korea\
\and
SNU Astronomy Research Center, Seoul National University, 1 Gwanak-ro, Gwanak-gu, Seoul 08826, Republic of Korea\
\and
Australian Astronomical Optics - Macquarie University, 105 Delhi Road, North Ryde, NSW 2113, Australia\
\and
Escuela de Ingeniería, Universidad Central de Chile, Avenida Francisco de Aguirre 0405, 171-0614 La Serena, Coquimbo, Chile\
\and
Steward Observatory, University of Arizona, 933 North Cherry Avenue, Tucson, AZ 85721, USA\
}

%\date{Received September 15, 1996; accepted March 16, 1997}
\date{Received XXX; accepted XXX}

% \abstract{}{}{}{}{} 
% 5 {} token are mandatory
 
  \abstract
  % context heading (optional)
  % {} leave it empty if necessary  
   {}
  % aims heading (mandatory)
   {We investigate if systems of multiple Lyman-alpha emitters (LAEs) can serve as a proxy for dark matter halo mass, assess how their radiative properties relate to the underlying halo conditions, and explore the physics of star formation activity in LAEs and its relation to possible physically related companions.}
  % methods heading (mandatory)
   {We use data from the One-hundred-deg$^2$ DECam Imaging in Narrowbands (ODIN) survey, which targets LAEs in three narrow redshift slices. We identify physically associated LAE multiples in the COSMOS field at $z = 2.4$, $z = 3.1$, and $z=4.5$, and use a mock catalog from the IllustrisTNG100 simulation to assess the completeness and contamination affecting the resulting sample of LAE multiples. We then study their statistical and radiative properties as a function of multiplicity, where we adopt the term multiplicity to refer to the number of physically associated LAEs.  
   }
  % results heading (mandatory)
   {We find a strong correlation between LAE multiplicity and host halo mass in the mocks, with higher multiplicity systems preferentially occupying more massive halos. In both ODIN and mock, the mean Ly$\alpha$ luminosity increases with multiplicity, while the mean UV luminosity shows weaker trends. Using the mock, we find that LAEs are unbiased tracers of the total halo UV luminosity.  In ODIN and in the mock, halo-wide LAE surface brightness densities in Ly$\alpha$ and UV increase with multiplicity, reflecting more compact and actively star-forming environments, particularly at $z = 3.1$. The close agreement between the model and ODIN-COSMOS observations supports the validity of the Ly$\alpha$ emission model in capturing key physical processes in LAE environments. Finally, a subhalo-based perturbation induced star formation model reproduces the minimum subhalo mass distribution in simulations at $z=2.4$, suggesting that local perturbations—rather than the presence of LAE companions—drive star formation activity in these systems. For the higher redshift samples, neighbor perturbations do not seem to be the main driver that triggers star formation.}
  % conclusions heading (optional), leave it empty if necessary 
   {}

   \keywords{Galaxies: high-redshift --
                Galaxies: halos --
                Galaxies: groups: general
               }

   \maketitle

%-------------------------------------------------------------------

\section{Introduction}

The high-redshift universe, a period of rapid galaxy formation, provides a unique laboratory for studying galaxy evolution. Lyman-alpha emitting galaxies (LAEs), characterized by strong emission in the Ly$\alpha$ line, are powerful probes of this era. Their properties, such as their high star formation rate and low metallicity, offer insights into the early stages of galaxy formation and the physical processes that drive their evolution (see \citealt{Ouchi2020}, for a review). 
LAEs are typically young, star-forming, low-mass systems with minimal dust and metal content \citep[e.g.,][]{Gawiser2007, Guaita2010, Ono2010}. 
They exhibit strong Ly$\alpha$ emission due to hydrogen recombination in their interstellar media (ISM), driven by active star formation or the presence of an active galactic nucleus (AGN) \citep[e.g.,][]{Kunth1998, Konno2016}.  
Although the mechanism behind the Ly$\alpha$ emission is still not fully understood, it originates at a rest-frame wavelength of $121.6$ $\mathrm{nm}$ and is redshifted into the optical range at redshifts $2 \lesssim z \lesssim 5$, making LAEs observable from the ground.
While LAEs generally are dust poor, low stellar mass galaxies (M$_* \leq 10^9$ M$_{\odot}$, \citealt[e.g.,][]{Lai2008, Ono2010, Kusakabe2018}); a small fraction may include high stellar mass, dusty galaxies. 

The complex radiative transfer processes undergone by Ly$\alpha$ photons make it difficult to find correlations between Ly$\alpha$ luminosity and other physical properties. However, some studies have shown a clear correlation between Ly$\alpha$ and UV continuum luminosities and halo mass using observational data over the redshift range $2.5 < z < 6$ \citep{Khostovan2019, Herrero-Alonso2023}, confirming previous tentative trends \citep{Ouchi2003, Kusakabe2018}.

Several studies show that LAEs are reliable tracers of the matter distribution, making them ideal for studying the large-scale structure of the universe \citep[e.g.,][]{Ouchi2020,Huang2022,Im2024}. From clustering measurements, \citet{Herrero-Alonso2023} find that the typical halo mass increases from $\log(M_h / [h^{-1} M_\odot]) \approx 10.0$ to $\approx 11.43$ between Ly$\alpha$ luminosities of $10^{40.97}$ and $10^{42.53}$ erg s$^{-1}$, while \citet{Khostovan2019} report an increase from $\log(M_h / [h^{-1} M_\odot]) \approx 9.75$ to $\approx 12.8$ over the range $10^{41.7}$ to $10^{43.6}$ erg s$^{-1}$.
These results fit well within the widely assumed framework in which star-forming galaxies that reside in more massive halos have higher star formation rates and therefore exhibit more luminous nebular emission lines \citep[e.g.,][]{Kusakabe2018,Jun2025}. This suggests that Ly$\alpha$ luminosity may also serve as an indirect tracer of strong star formation activity and dark-matter halo mass.  The study of this relation could help to understand the star formation processes at high redshifts.

Depending on the selection process, the number of galaxies in a halo can be related to the halo mass in a monotonic way (e.g. \citealt{zehavi}).
Halo occupation distribution (HOD) analyses of LAEs suggest that the most common scenario is the presence of a single detected LAE per halo, with a small but robust satellite fraction \citep{Herrero-Alonso2023}. In principle, if multiple LAEs are present within halos, even if these are rare, LAE occupancy could be used to select halos of different masses, since occupancy in halo models typically increases monotonically with halo mass \citep[e.g.][]{BerlindWeinberg2002,Yang2008}.  A sample of multiples from LAEs could then be used to study the relation between halo mass and the physics of star formation. For this purpose, a large and reliable sample of LAEs over a wide area of the sky is required so as to build a sample of LAE multiples with significant numbers.

For the study of LAEs with large, uniform samples and at different cosmic epochs, an ideal dataset is the One-hundred-deg$^2$ DECam Imaging in Narrowbands (\citealp[ODIN,][]{Lee2024}) survey. This program aims to detect more than 100,000 LAEs over large contiguous areas using narrow-band filters, allowing one to map the universe at redshifts of 2.4, 3.1, and 4.5.
Preliminary samples from several ODIN fields have already been utilized in various recent studies, demonstrating their significant utility in both protocluster identification and clustering analyses.
For instance, \cite{Ramakrishnan2025} conducted a systematic search for protoclusters at Cosmic Noon using ODIN LAEs as tracers and identified 150 protocluster candidates at $z = 2.4$ and $z = 3.1$ across 13.9 deg$^2$, showing a high clustering signal and providing insights into the redshift evolution of cluster progenitors.
In a complementary study, \cite{Andrews2025} analyzed galaxies in simulated protocluster environments at Cosmic Noon using IllustrisTNG. By modeling the LAE population, they provided expectations for the ODIN survey and reinforced the idea that the large-scale environment significantly impacts galaxy evolution.
\cite{Herrera2025} studied the clustering of over 14,000 ODIN LAEs in 9 deg$^2$ at three redshifts, finding that LAEs typically occupy a small fraction ($3 - 5\%$) of dark matter halos and appear unusually luminous relative to their host halo masses.
\cite{Firestone2025}, in turn, investigated the star formation histories of 74 LAEs from ODIN, finding that $\sim 95\%$ are undergoing their largest burst of star formation to date, supporting the idea that LAEs are in a formative stage and differ from other star-forming galaxy populations, such as Lyman Break Galaxies (LBGs).

In this work, we use the ODIN data to propose LAE multiplicity as an alternative proxy for halo masses, based on the hypothesis that, in principle, a higher number of LAEs implies a greater mass. This will allow us to study how star formation depends on local halo environment. We will use mock catalogs to assess the completeness and contamination of our multiples. The mock catalog will be built using the IllustrisTNG100 simulation, coupled with the \cite{DijkstraWyithe2012} model for Ly$\alpha$ emission, ensuring consistent number densities as well as Ly$\alpha$, rest-frame UV luminosity functions and Ly$\alpha$ equivalent width distributions. With this mock sample, we will test whether the multiplicity hypothesis is indeed related to halo mass.
To find multiples, we base our approach on a previously developed method for identifying minor galaxy systems in the local universe \citep{Cerdosino2024}, adapted here to ODIN LAEs. With this new independent method, we aim to confirm that Ly$\alpha$ luminosity is higher for halos with larger masses M$_{200}$. Additionally, our analysis will allow us to study how the total luminosity (in both UV and Ly$\alpha$) and the virial density of these quantities depend on multiplicity and halo mass. This approach provides an opportunity to investigate the mechanisms that trigger star formation in intermediate stellar mass galaxies at high redshift, which in turn is the main process responsible for producing detectable Ly$\alpha$ emission. 
Additionally, we expect to test the validity of previous empirical models that relate the physical properties of galaxies to Ly$\alpha$ luminosity \citep[see, e.g.,][]{Weinberger2019} and explore the effect of halo mass on the star formation activity in LAEs.

This paper is organized as follows. Section \ref{sec:data} presents the ODIN dataset we use in this work and the mock catalog we use to interpret the results. In Section \ref{sec:lae_samples} we carefully define the behavior of LAEs in the mock and the LAE samples. Section \ref{sec:optimization} describes the method to obtain associations of multiple LAEs in both data and simulation, including the optimization process and the resulting multiples. Section \ref{sec:results} shows the results, including their dependence on Ly$\alpha$ and UV luminosities, followed by a discussion about the potential of LAEs to trace the group- or halo-wide projected Ly$\alpha$ and UV luminosity densities and the perturbation-induced star formation. Finally, we present our conclusions in Section \ref{sec:conclusions}.
Throughout this paper, we assume the standard $\Lambda$CDM cosmology of \citet{Planck2016} with $\Omega_{\rm m}$ = 0.3089, $\Omega_{\Lambda}$ = 0.6911 and $h$= 0.6774.

\section{Data} \label{sec:data}

\subsection{Observational data} \label{sec:odin_data}

The ODIN survey utilizes the Dark Energy Camera \citep[DECam,][]{Flaugher2015} on the Víctor M. Blanco 4m telescope at the Cerro Tololo Inter-American Observatory (CTIO) in Chile. This project employs three custom-designed narrowband filters with central wavelengths at 419 nm ($N419$), 501 nm ($N501$), and 675 nm ($N673$) to target LAE candidates at redshifts $2.4$, $3.1$, and $4.5$, respectively. The survey aims to discover an unprecedented number of LAEs across seven deep and wide fields, covering a final area of $\sim 100$ deg$^2$, which will allow ODIN to construct samples of the galaxy population in three distinct epochs in cosmological history.    
Here, we use the observations at redshifts $2.4$, $3.1$, and $4.5$ of the COSMOS field ($\sim 9$ deg$^2$).
   
The method used to select LAEs is described in \citet{Firestone2024}; here, we summarize the selection process.
As a first step, all data were corrected for galactic dust and aperture corrections to estimate flux densities. Additionally, starmasking was applied to remove data contaminated by saturated stars and the effects of pixel oversaturation in the camera. Further cuts were also applied to improve the quality of the data.
To select LAE candidates, the ODIN team developed a new "hybrid-weighted double-broadband continuum estimation technique". Their method estimates the expected continuum emission at the narrowband's effective wavelength by using data from two nearby broadband filters. 
Then, to infer the presence of an emission line at the redshifted Ly$\alpha$ wavelength, they look for excess flux density in the narrowband with respect to the estimated continuum. If this excess is higher than the minimum value of the rest-frame Ly$\alpha$ equivalent width (REW), set at 20 \r{A}, then the source is considered an LAE candidate.
The ODIN team applied this technique to select LAE candidates at the three redshifts, using broadband data from the Hyper Suprime-Cam Subaru Strategic Program (HSC-SSP) \citep{Kawanomoto2018,Aihara2019} and narrowband data collected with DECam. 
They used the $N419$, $r$, and $g$ bands for the $z = 2.4$ LAE selection; the $N501$, $g$, and $r$ bands for $z = 3.1$; and the $N673$, $g$, and $i$ bands for $z = 4.5$.
Finally, they developed a method to identify the largest known sources of contamination and to enhance the purity of the LAE samples.

In addition, the ODIN group derived the REWs for the LAE samples (\citealt{Firestone2024}), following the methodologies of \cite{Venemans2005} and \cite{Guaita2010}. They also estimated the Ly$\alpha$ line fluxes, which we used to compute the Ly$\alpha$ luminosities.
We then derived the UV continuum magnitudes, M$_\text{UV}$, from the UV continuum flux densities $f_{\text{cont, Ly}\alpha}$ at the wavelength of the Ly$\alpha$ line. This quantity can be estimated from the observed measurements of the Ly$\alpha$ line flux $F_{\text{line, Ly}\alpha}$ and the REW mentioned above, following the equation: 

\begin{equation}
    \text{REW} = \int_{\lambda_0}^{\lambda_1} \frac{ f_{\text{NB, Ly}\alpha} - f_{\text{cont, Ly}\alpha}} {f_{\text{cont, Ly}\alpha}} \, d\lambda \approx \frac{F_{\text{line, Ly}\alpha}}{f_{\text{cont, Ly}\alpha}},
\end{equation}
where $\lambda_0$ and $\lambda_1$ define the integration range, which is the width of the Ly$\alpha$ line, and $f_{\text{NB, Ly}\alpha}$ is the flux density of the narrowband. This approximation is valid if $f_{\text{cont, Ly}\alpha} << f_{\text{NB, Ly}\alpha} $. We then apply a correction to shift the magnitudes to 1600 $\text{\r{A}}$, to match the rest-frame wavelength of the mock magnitudes (see Section \ref{sec:mock_data}).

\begin{figure*}[h!]
\centering
\includegraphics[width=0.95\textwidth]{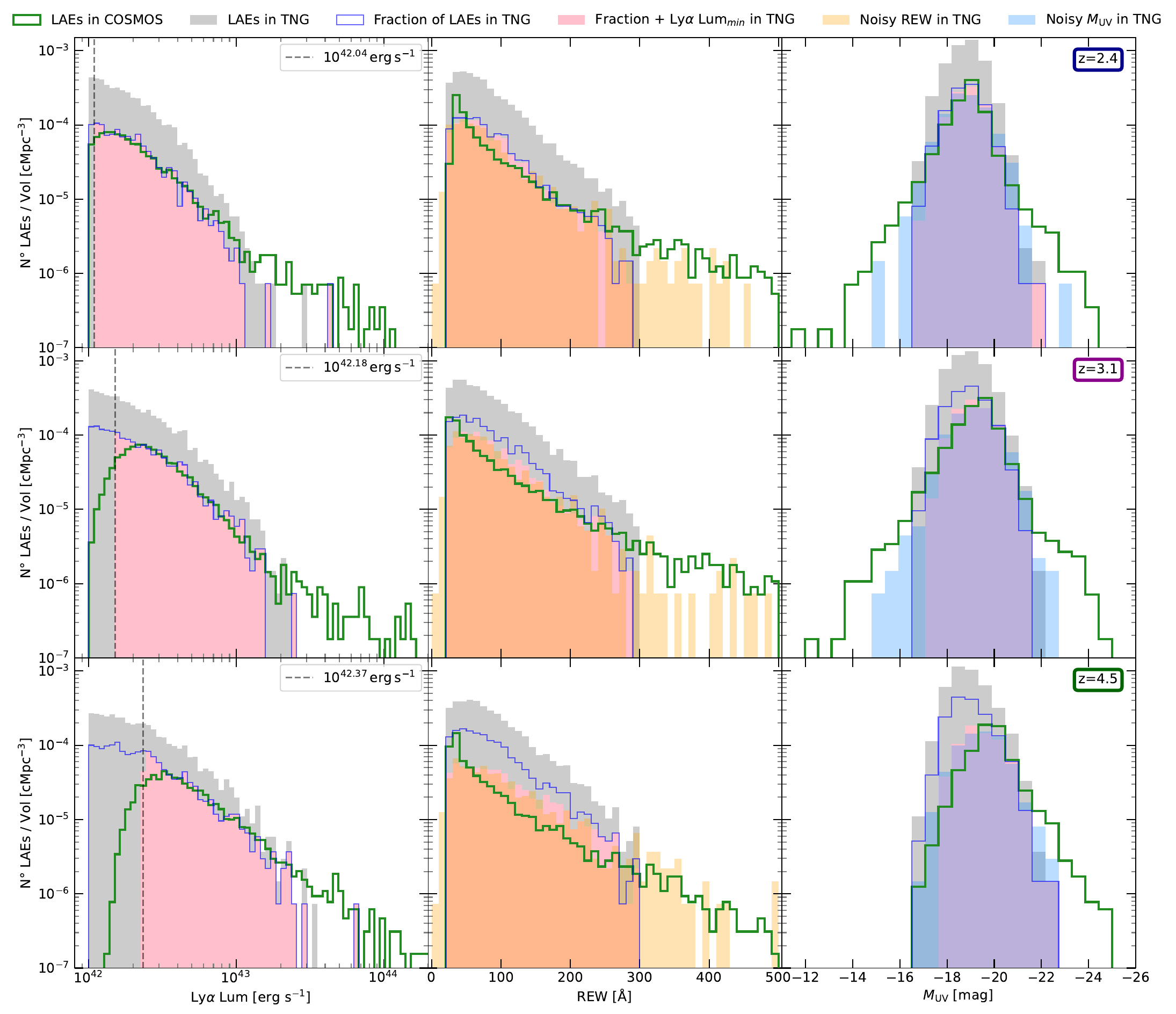} 
\caption{Ly$\alpha$ luminosity (\textit{left panels}), rest-frame equivalent width (REW, \textit{middle panels}) and UV absolute magnitude (M$_\text{UV}$, \textit{right panels}) distributions per unit volume for redshifts $z=2.4$ (\textit{top panels}), $z=3.1$ (\textit{middle panels}) and $z=4.5$ (\textit{bottom panels}).
The green solid line shows the ODIN-COSMOS LAE sample (Ly$\alpha$ Lum $\geq$ 10$^{42}$ erg s$^{-1}$ and REW $\geq$ 20 \r{A}), while the gray shaded area shows the same sample in the TNG100 simulation.
The blue solid line shows a randomly selected subset representing 25$\%$, 33$\%$, and 40$\%$ of the LAEs in TNG100 at $z=2.4$, $3.1$, and $4.5$, respectively. 
The pink shaded areas show the latter samples applying a minimum luminosity cut of $10^{42.04}$, $10^{42.18}$, and $10^{42.37}$ erg s$^{-1}$ (vertical dashed line shown in the left panels), respectively.
The orange shaded areas show the REW distributions noisified by photometric uncertainties, while the sky blue shaded regions show the corresponding noisified M$_\text{UV}$ distributions (see text).
}
\label{fig:lya_mag_rew}
\end{figure*}

To define our LAE sample, we apply selection cuts: Ly$\alpha$ luminosities above 10$^{42}$ erg s$^{-1}$ and REWs exceeding 20 \r{A} \citep[see][]{Firestone2024}. These criteria result in samples of 5714, 5710, and 4101 LAEs at redshifts $2.4$, $3.1$, and $4.5$, respectively. 
The solid green lines in Figure \ref{fig:lya_mag_rew} show the volume-normalized distributions of Ly$\alpha$ luminosity, REW, and UV absolute magnitudes for the ODIN-COSMOS LAE samples at the three redshifts. This criterion, in addition to being consistent with those adopted by the rest of the ODIN collaboration, is also aligned with the fact that selecting LAEs with relatively high luminosities (Ly$\alpha$ Lum $\approx 10^{42}$ erg s$^{-1}$) increases the likelihood of probing higher density regions \citep{Herrero-Alonso2023}, which is particularly useful for our sample of multiples.

\subsection{Mock data} \label{sec:mock_data}

We model LAEs following \cite{Andrews2025}, who implement a model for LAE galaxies in \textit{The Next Generation} Illustris \citep[IllustrisTNG,][]{Nelson2019,Pillepich2018}. 
The IllustrisTNG is a set of magneto-hydrodynamical cosmological simulations run with the \textsc{Arepo} moving mesh code \citep{Springel2010}. These simulations include sub-grid models that account for star formation, radiative metal-line gas cooling, chemical enrichment from supernovae types II and Ia, as well as asymptotic giant branch stars, stellar feedback, and supermassive black holes growth and kinetic feedback. 
The differing particle resolutions and box sizes render the IllustrisTNG simulations complementary, allowing them to be particularly useful for distinct analyzes depending on the specific goals of each study.
In particular, IllustrisTNG100-1 (hereafter TNG100), is a box with a side length of $110.7$~cMpc run using initial conditions with $1820^3$ dark matter particles of mass $7.5 \times 10^6 {\rm M_{\odot}}$ and $1820^3$ gas cells with a mass of $1.4 \times 10^6 {\rm M_{\odot}}$.  %TNG100 offers a  resolution that makes it adequate to simulate our LAE sample.   

The LAE model follows the prescription given in \cite{DijkstraWyithe2012} and is further developed by \cite{Gronke2015,Weinberger2019}. It assigns the Ly$\alpha$ luminosity, $L_{\text{Ly}\alpha}$, to galaxies based on their UV luminosity density, $L_{\text{UV}, \nu}$, (in our case derived from \citealt{Vogelsberger2020}, model c), and a randomly drawn REW from a probability distribution function, $P(\rm REW| \rm M_{\rm UV})$, according to the following equation:

\begin{equation}
    L_{\text{Ly}\alpha} = \frac{\nu_{\alpha}}{\lambda_{\alpha}} \left( \frac{\lambda_{\text{UV}}}{\lambda_{\alpha}} \right)^{-\beta - 2} \times \text{REW} \times L_{\text{UV}, \nu},
    \label{eq:lumlya}
\end{equation}
where $\nu_{\alpha} = 2.47 \times 10^{15} \, \text{Hz}$, $\lambda_{\alpha} = 1216 \, \text{\r{A}}$, $\beta = -1.7$ is the UV spectral slope, and $\lambda_{\text{UV}} = 1600 \, \text{\r{A}}$ denotes the rest-frame wavelength at which the UV continuum flux density was measured.
The probability distribution $P(\rm REW| \rm M_{\rm UV})$ follows the form:
\begin{equation}
%\boldsymbol{P}(\boldsymbol{\mathrm{REW}}\,|\,\mathbf{M}_{\mathrm{UV}}) = \boldsymbol{N} \exp\left(-\frac{\boldsymbol{\mathrm{REW}}}{\boldsymbol{\mathrm{REW}}_c(\mathbf{M}_{\mathrm{UV}})}\right),
P(\rm REW \,|\, \rm M_{\rm UV}) = N \exp\left( -\frac{\rm REW}{\rm REW_{\rm c}(M_{\rm UV})} \right)
\end{equation}
where $\rm REW_{\rm c}$ is a characteristic REW that depends on the UV magnitude, and $N$ is a normalization constant.
For more details, see \cite{Andrews2025}.

In this work, we use the model applied to TNG100 with a lower stellar mass of M$_*>10^{8.14}$ M$_{\odot}$ to ensure completeness of the modeled galaxy sample in TNG100. The resulting mock sample has the same minimum halo mass estimated from ODIN observations \citep{Herrera2025}, as well as similar clustering and bias obtained assuming a uniform HOD.

\section{LAE samples} \label{sec:lae_samples}

As with the ODIN data (Section \ref{sec:odin_data}), mock galaxies are classified as LAEs if they satisfy the same conditions, Ly$\alpha$ luminosity $\geq$ 10$^{42}$ erg s$^{-1}$ and REW $\geq$ 20 \r{A} (gray shaded area in Fig. \ref{fig:lya_mag_rew}). 
However, we noticed that the number density of LAEs in TNG100 is higher than in ODIN-COSMOS, with values of $(4$–$5) \times 10^{-3}$ Mpc$^{-3}$ and $1 \times 10^{-3} - 6 \times 10^{-4}$ Mpc$^{-3}$, respectively. This discrepancy between the observed and predicted number densities is expected as described by \cite{Andrews2025} and \cite{DijkstraWyithe2012} through a normalization factor $F$. The mismatch arises because the model does not fully capture the complexities of Ly$\alpha$ radiative transfer, including the internal structure and geometry of dust and gas, as well as the star formation histories of LAEs, which can add a stochastic component to LAE detection. 
Therefore, only a fraction of galaxies that satisfy the conditions to be detected as LAE actually make it into the observed LAE sample. This is likely due in part to viewing-angle effects, such as Ly$\alpha$ emission being unobscured only along certain lines of sight.
Consequently, we randomly select a fraction of the simulated LAEs to match the observed amplitude of the distribution of ODIN LAE luminosities. For $z = 2.4$, we used 25$\%$ of the simulated sample, for $z = 3.1$, we used 33$\%$, and for $z=4.5$, we used 40$\%$. These subsamples are shown as solid blue lines in  Figure \ref{fig:lya_mag_rew}.
It is important to note that this random subsampling alone does not guarantee that the space densities of the mock and observed samples will match. To achieve a precise correspondence, an additional criterion based on luminosity must be applied, as detailed below.

We also noticed an incompleteness in the observed Ly$\alpha$ luminosity distribution of ODIN-COSMOS LAEs compared to the mocks (see the left panels of Fig. \ref{fig:lya_mag_rew}), which can be attributed to limitations in the observational setup. This incompleteness is evident at the faint end of the distribution, where the number of LAEs is underestimated in the observational data. As a result, a clear drop in source counts is visible at lower luminosities.
For that reason, we applied a minimum luminosity cut to the TNG100 LAEs to remove a percentage of the lowest-luminosity sources that are less complete in the observations. The cut was chosen to produce a volumetric number density that matches the number densities of the ODIN samples. This corresponds to luminosity thresholds of 10$^{42.04}$ erg s$^{-1}$ for $z = 2.4$, 10$^{42.18}$ erg s$^{-1}$ for $z = 3.1$ and 10$^{42.37}$ erg s$^{-1}$ for $z = 4.5$ (vertical dashed lines in the left panels of Fig. \ref{fig:lya_mag_rew}). These subsamples are shown in pink shaded histograms and are the TNG100 samples that we will consider to identify the LAE multiples in later sections.  

It is important to note that, to make the mock catalogs comparable to the line-of-sight “thickness” of the observational data, we constructed 2D projections of the simulation over line-of-sight depths of $\sim 74$ cMpc at $z=2.4$, $\sim 60$ cMpc at $z=3.1$, and $\sim 50$ cMpc at $z=4.5$ \citep{Lee2024}. We added together different projection axes and several different random seeds when selecting the LAE fractions so as to match the comoving volumes of the observations at each redshift. 

The left panels of Fig. \ref{fig:lya_mag_rew} also show that ODIN-COSMOS LAEs exhibit a high-luminosity tail, which is typically attributed to the presence of AGNs \citep[e.g.,][]{Konno2016,Sobral2018}. This feature is not reproduced by the model as it does not include the AGN emission.

In addition, it is important to note that the model only reproduces systems with REWs up to 300 \r{A} (middle panels of Fig. \ref{fig:lya_mag_rew}). Extremely high REW values (greater than 240 \r{A}) observed in the ODIN-COSMOS sample may represent genuine signals \citep[e.g.,][]{Inoue2011,Kashikawa2012}, but they could also result from spurious measurements introduced by low signal-to-noise ratios in photometry \citep{Firestone2024}. 
This discrepancy at the high end of the REW distribution appears to be a systematic effect introduced when real photometry is used. Specifically, the observed excess of high REW values could be driven by sources with large photometric uncertainties in the broadband filters used to estimate the continuum. These uncertainties would amplify the derived REWs, particularly for objects with faint continuum detections. 
To test this, we construct a flat (top-hat) input REW distribution sampled from real ODIN data with associated photometric errors. By perturbing the narrowband and double broadband magnitudes within their 1$\sigma$ uncertainties and recalculating the REW, we reproduce an artificial tail of high REW values. We find that this tail persists even when only the double broadband is noisified, which suggests that the broadband errors dominate the effect. This is consistent with expectations, as LAEs with large REWs are often faint in the continuum, leading to significant broadband uncertainties.
To account for this in the mock sample, we applied a “noisification” procedure to the REW values derived from the model. The resulting mock sample, including these photometric uncertainties, is shown as the orange shaded distribution in Fig. \ref{fig:lya_mag_rew}. In this way, the tail of the high REWs is reproduced and no additional REW cut is required.

Regarding the UV magnitude distributions (right panels in Fig. \ref{fig:lya_mag_rew}), the observations display a greater dispersion toward both brighter and fainter values. %At $z=4.5$, the central peak of the distribution aligns well with that of the simulation samples once the stochastic fraction and the minimum LAE luminosity are adjusted for ODIN incompleteness. However, at lower redshifts, the mock samples exhibit a shift toward brighter magnitudes. 
At higher redshifts the ODIN-COSMOS distributions align well with those of the simulation samples once the stochastic fraction and the minimum LAE luminosity are adjusted for ODIN incompleteness.
Once again, we aim to assess the effect introduced by using real broadband photometry in the continuum estimations. To do so, we compute the UV magnitudes from the noisified REW distributions, effectively assuming this noise is mostly due to the continuum flux estimate, using Equation~\ref{eq:lumlya}. The resulting distributions are shown as sky blue shaded regions and exhibit increased scatter, which tends to better reproduce the observed M$_\mathrm{UV}$.

\subsection{Analysis of LAE host halos in IllustrisTNG}

LAEs are crucial for understanding galaxy formation at high redshifts. However, their representativeness within the overall galaxy population remains uncertain. In TNG100, only 15$\%$, 17$\%$, and 18$\%$ of the total galaxy population are classified as LAEs at $z = 2.4$, $z = 3.1$, and $z=4.5$, respectively, according to our LAE selection criteria (i.e., Ly$\alpha$ Lum $\geq$ the adopted threshold and REW $\geq$ 20 \r{A}). These fractions suggest that LAEs may not fully represent the diversity of high-redshift galaxy populations. 
In Appendix \ref{appen:frac_of_laes} we further analyze the behavior of LAEs and their fraction relative to the total galaxy population, where we find that LAEs preferentially reside in halos of intermediate mass, as well as in galaxies with intermediate stellar masses and moderate star formation rates, consistently across the three redshifts.

\begin{figure}
\centering
\includegraphics[width=0.4\textwidth]{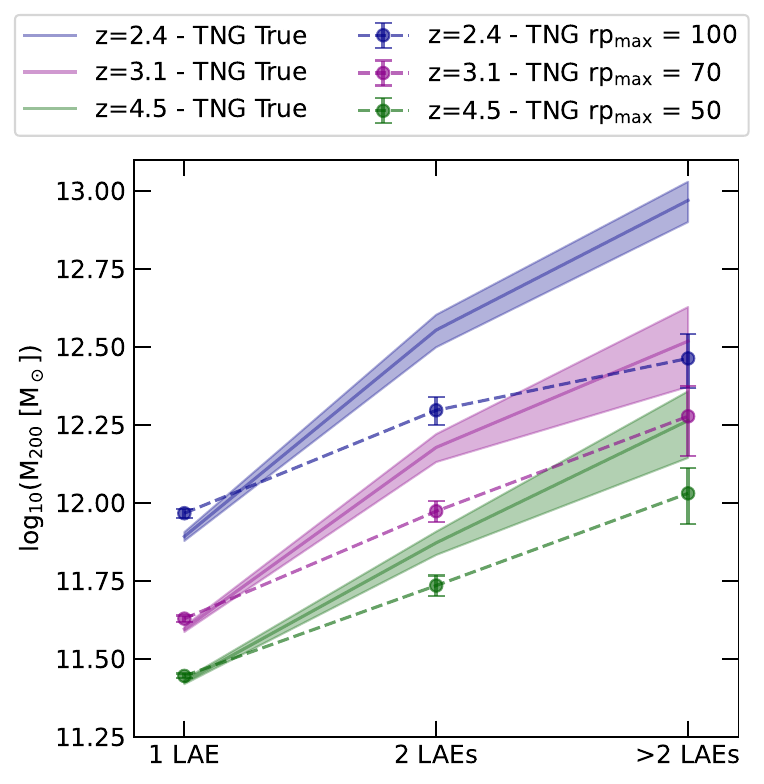} 
    \caption{Mean halo mass as a function of the number of LAEs within the halo at $z=2.4$ (blue), $z=3.1$ (magenta) and $z=4.5$ (green). The TNG100 true multiples are shown with solid lines, while the TNG100 multiples identified using projected distances of $rp_{max} = 100$ pkpc at $z=2.4$, $70$ pkpc at $z=3.1$ and 50 pkpc at $z=4.5$ are displayed in dashed lines. Shaded regions and error bars show the error of the mean. As the number of LAEs in a halo increases, the true mean halo mass also increases, suggesting that the number of physically associated LAEs can serve as a proxy of halo mass. The identified multiples follow the underlying trend more closely at higher redshift.}
    \label{fig:masas_nlae}
\end{figure}

Depending on the galaxy selection, more massive halos can host a greater number of galaxies. Here, we will explore whether this is the case for the number of LAEs using a simulated catalog. Furthermore, we can make further tests, such as whether higher-mass halos contain more luminous LAEs, and whether the virial UV and Ly$\alpha$ luminosity densities depart from a constant value, as is the case of the virial mass density. 
Based on the analysis in Appendix \ref{appen:frac_of_laes}, we expect that only about $10 - 20 \%$ of massive halo populations are represented by LAEs. Therefore, we investigate whether multiple LAEs can be found within the same halo in our TNG100 samples. To do this, we examine the relationship between the halo mass M$_{200}$ and the number of LAEs each halo contains. 

The solid lines in Figure \ref{fig:masas_nlae} show the mean mass of the modeled halos at the three redshifts as a function of the number of LAEs they contain. The LAE samples used here are defined in Section \ref{sec:lae_samples}, based on the corresponding 2D projections.
From this analysis, we find that $\sim 97-98\%$ of LAEs are the only LAE within their host halo, which typically has a mean mass of $\rm M_{200} \sim 10^{11.4} - 10^{11.9} \rm M_{\odot}$, increasing toward lower redshifts. In other words, if we consider a survey that includes exclusively LAEs, we may safely assume that most of these galaxies are isolated. However, there are $\sim 2-3 \%$ of halos that host two LAEs, and about $\sim 0.1-0.2 \%$ that contain three or more. This is something we will take into account to limit our selection of LAE multiples.

In addition, Figure \ref{fig:masas_nlae} reveals that as the number of LAEs within a halo increases, the halo mass also increases. %, suggesting that these LAEs have more chance to be disturbed by nearby neighbors than the isolated ones. 
The mean $\rm M_{200}$ increases from $10^{11.9}$ to $10^{13} \rm M_{\odot}$ at $z=2.4$, from $10^{11.6}$ to $10^{12.5} \rm M_{\odot}$ at $z=3.1$, and from $10^{11.4}$ to $10^{12.26} \rm M_{\odot}$ at $z=4.5$ with increasing the number of LAEs from 1 to >2.

Building on this, in the following Section we search for LAE multiples in the mock catalog and in ODIN. A mass dependence on LAE multiplicity implies that multiple LAEs are more likely to be influenced by interactions with nearby galaxies, potentially leading to observable differences in key physical properties such as stellar mass and star formation rate.

\section{Optimizing the identification of galaxy multiples} \label{sec:optimization}

We now proceed to identify LAE multiples with the aim of identifying physically associated systems of varying masses and properties. To do so, we take advantage of the fact that our LAEs are located at similar redshifts due to the narrow width of the ODIN custom filters. The ODIN survey provides redshift precision of $\Delta z \sim 0.03 - 0.04$, which is significantly better than the precision typically achievable with photometric redshifts derived from broadband filters ($\Delta z \sim 0.1 - 0.4$ at $z \gtrsim 2$, e.g. \citealt{Shi2020,Huang2022}). Specifically, the FWHM of the $N419$, $N501$ and $N673$ filters corresponds to $\Delta z = 0.061$, $0.062$ and $0.082$, respectively, in redshift space \citep{Lee2024}. 

Several methods for identifying galaxy systems have been developed over the years \citep[e.g.,][]{Lambas2003,OMill2012}. In this work, we adapt a technique originally applied to minor galaxy systems at low redshift \citep{Cerdosino2024}, modifying it for application to our high-redshift LAE samples.
Robust system identification requires a methodology that carefully balances both the purity and completeness of detected systems. These two aspects are best evaluated using mock catalogs, which allow for precise control over the true physical associations between galaxies \citep[e.g.,][]{Rodriguez2020}. For this purpose, we use our mock LAE catalog (described in Section \ref{sec:lae_samples}) which provides both the projected positions and host halo information for each LAE.

We compute the projected distances between LAEs and impose a minimum separation of 10 pkpc. This lower limit is motivated by the observational angular resolution of ODIN, which corresponds to $\sim 1.2$, $\sim1.3$, and $\sim1.5$ arcsec at $z = 2.4$, $z = 3.1$, and $z=4.5$, respectively.
In cases where two mock LAEs are closer than this limit, we retain only the one with the higher Ly$\alpha$ luminosity. This minimum distance is consistent with definitions commonly adopted in other works on galaxy systems at high redshift \citep[e.g.,][]{Matthee2023}. While this choice represents a simplifying assumption, favoring the brighter source, in real observations such close pairs would be merged. Additionally, it is likely that not all the luminosity of the fainter unresolved source would be accounted for in the merged one. It is worth noting, however, that this situation accounts for less than $0.5\%$ of galaxies in all cases.
We also estimate that the impact of peculiar velocities is negligible, as they are on the order of a few hundred km s$^{-1}$, representing only a small fraction of the survey depth.

To group the LAEs, we search for neighboring LAEs within a maximum projected separation $rp_{max}$ around each one and use a graph-based method to connect them. 
In this approach, pairs of galaxies separated by less than $rp_{max}$ are linked without repeating galaxies, and the connected components of the resulting graph are interpreted as galaxy associations. Each association corresponds to a set of galaxies that are spatially connected to one another within the adopted distance threshold.
We then determine the optimal value of $rp_{max}$ that provides the best balance between contamination and completeness.
We use two definitions to classify our LAE multiples. The first, "TNG100 true", refers to cases where all LAEs are in the same halo in projection, without applying a maximum projected separation. The second, "TNG100 identified", includes multiples identified in projection after imposing a maximum projected distance cut-off. This latter definition is intended to mimic the identification applied to observations.

To evaluate our selection method we use two definitions: contamination and fraction of halos recovered (completeness). A detailed description of these definitions, together with the methodology used to compute them, is provided in Appendix \ref{appen:test_cyc}.
To ensure that contamination remains reasonably low while maintaining a substantial fraction of recovered halos, we chose a maximum projected distance of 100 pkpc for $z=2.4$, 70 pkpc for $z=3.1$, and 50 pkpc for $z=4.5$. 
This choice results in a recovered halo fraction of approximately $54\%$, $51 \%$, and $54 \%$ with a contamination of about $55\%$, $58 \%$, and $49 \%$ at $z=2.4$, $z=3.1$, and $z=4.5$, respectively.  
These values were selected not only to balance completeness and contamination, but also to optimize the correlation between halo mass and the number of LAEs in the identified multiples (see next subsection).

These values of contamination are higher than what can be obtained with other low-$z$ surveys \citep{Cerdosino2024}. Our current ODIN LAEs still contain limited information available only including UV luminosities, Ly$\alpha$ luminosities, and equivalent widths, with a fixed photometric redshift error. We carefully searched for ways to improve contamination using the true and falsely identified LAE associations in the TNG100 mocks and found no differences in the distributions of relative UV or Ly$\alpha$ luminosities that could have been used to help reduce contamination. As the survey progresses, the addition of physical properties estimated for the LAE candidates may help to improve future searches for multiples in ODIN.  However, this first sample provides interesting insights into LAE groups and their physical properties, as we show below.

\subsection{Multiples of LAEs}

We use the TNG100 identified multiples to assess whether the median mass of the halos increases according to the number of associated LAEs, as previously observed for the TNG100 trues in Figure \ref{fig:masas_nlae}.
The blue dashed line in that figure shows the multiples identified with a $rp_{max}$ of 100 pkpc at $z=2.4$, the magenta dashed line represents those identified with 70 pkpc at $z=3.1$, and the green dashed line those with 50 pkpc at $z=4.5$.
The mean halo masses recovered from the TNG100-identified multiples follow the overall trend observed in the TNG100-true cases, with the correlation between halo mass and number of LAEs being more evident at higher redshifts. 
We observe a systematic effect that tends to lower the mean halo masses of identified groups with higher LAE multiplicity. This bias is primarily driven by the incorrect inclusion of low-mass, isolated LAEs in the identification of multiples, which artificially lowers the recovered masses.

Using the same $rp_{max}$ values, we identified the multiples in the ODIN-COSMOS LAE sample. % (Ly$\alpha$ lum $\geq 10^{42}$ erg s$^{-1}$ and REW $\geq 20$ \r{A}).
Table \ref{table:numbers} presents the number of identified multiples for both the mock and observational datasets, as well as the percentage relative to the total number of LAEs in each sample.

\begin{table}
\caption{Number and percentage of TNG100 trues, TNG100 identified, and ODIN-COSMOS multiples, using $rp_{max} = 100$ pkpc at $z=2.4$, $rp_{max} = 70$ pkpc at $z=3.1$ and $rp_{max} = 50$ pkpc at $z=4.5$. They are classified by the number of LAEs they contain: 1, 2, 3 or more, and total.}             
\label{table:numbers}      
\centering           
\begin{tabular}{c c c c c c c}     
\hline\hline       

N° of  & \multicolumn{2}{c}{TNG True} & \multicolumn{2}{c}{TNG Identified} & \multicolumn{2}{c}{COSMOS} \\ 
LAEs & $\#$ & $\%$  &  $\#$ & $\%$  &  $\#$ & $\%$  \\ 

\hline      
   $z=2.4$ \\ %rp=100
   1         & 5204 & 96.71  & 5185 & 96.48  & 5489 & 98.02\\  
   2         & 166  & 3.08   & 180  & 3.35   & 108  & 1.93\\
   >2        & 11   & 0.20   & 9    & 0.17   & 3    & 0.05\\
   Total     & 5381 & 100    & 5374 & 100    & 5600 & 100\\

\hline      
   $z=3.1$\\ %rp=70
   1         & 6590 & 97.70  & 6567 & 97.50  & 5554 & 98.61\\
   2         & 143  & 2.12   & 160  & 2.37   & 78   & 1.38\\
   >2        & 12   & 0.18   & 8    & 0.12   & 0    & 0\\
   Total     & 6745 & 100    & 6735 & 100    & 5632 & 100\\

\hline      
   $z=4.5$ \\ %rp=70
   1         & 4644 & 97.89  & 4644 & 97.87  & 4037 & 99.24\\
   2         & 94   & 1.98   & 96   & 2.02   & 29   & 0.71\\
   >2        & 6    & 0.13   & 5    & 0.10   & 2    & 0.05\\
   Total     & 4744 & 100    & 4745 & 100    & 4068 & 100\\

\hline                  
\end{tabular}
\end{table}

The vast majority of LAEs in both the simulated and observational samples are isolated, with more than $96\%$ of sources appearing as single LAEs across the three redshifts. 
Only a small fraction are identified as multiples of 2 LAEs ($\sim 1 - 3\%$), and even fewer ($\leq 0.2\%$) as part of larger associations. 
These trends are consistent across the TNG100-true, TNG100-identified, and ODIN-COSMOS samples. However, ODIN-COSMOS shows a slightly higher fraction of isolated LAEs ($\sim 98-99\%$) and correspondingly lower percentage of higher-order multiples.
The similarity between the TNG100 true and TNG100 identified multiplicity distributions suggests that our graph-based method is effective in retrieving the overall statistics of LAE multiples. 
The close agreement between the multiplicity statistics in the observational and TNG100 samples is strongly dependent on the stochasticity introduced when constructing the mock catalog, where only a fraction of the simulated LAEs is randomly selected. Without this selection step, the simulation produces a significantly larger number of LAE multiples, which is inconsistent with observational findings. This can be taken as an additional indication that the Ly$\alpha$ luminosity model applied to TNG100 galaxies is able to capture quite successfully the physics and stochasticity behind this type of emission. The small remaining discrepancy in the ODIN-COSMOS sample, particularly for associations with more than two members, may be attributed to observational limitations, such as ODIN sensitivity thresholds or projection effects due to the filter’s depth along the line of sight.

As discussed in Section \ref{sec:lae_samples} regarding the potential presence of AGNs in our sample of LAEs, we further classified our ODIN-COSMOS multiples into two categories: those in which all LAEs have Ly$\alpha$ luminosities less than 10$^{43.5}$ erg s$^{-1}$, and those containing at least one LAE with a Ly$\alpha$ luminosity exceeding this threshold, which could indicate the presence of a potential AGN \citep{Konno2016,Sobral2018}. 
The majority of the multiples are composed of LAEs with lower luminosities, with only $\leq 2\%$ containing a possible AGN for the samples at redshifts 2.4 and 3.1. At $z=4.5$, this percentage is a bit higher ($3.4 \%$).
To compare these results with a matched random sample, we selected pairs and triples of isolated LAEs in higher proportions than those found in the identifications, choosing as many pairs and triples as possible from random realizations without replacement ($27\%$ for pairs and $18\%$ for triples). The random sample yielded similar percentages: at $z = 2.4$ and $z = 3.1$, less than $2\%$ of multiples have a LAE with luminosities above 10$^{43.5}$ erg s$^{-1}$, while slightly higher values ($3\text{--}5\%$) are found at $z = 4$.
Interestingly, none of the multiples consisting of more than 2 LAEs at $z=2.4$ includes a potential AGN. However, this is similar to the result of the random samples indicating no clear relation between AGN activity and LAE multiplicity, an assertion that can be revisited with the larger final ODIN samples. 
Note that this behavior is not typically expected based on studies of protoclusters. At redshifts between 2 and 6, AGNs are often found to be more prevalent in overdense environments, as shown in several works \citep[e.g.,][]{Lehmer2009, Casey2015, Shah2024}. This contrasts with the local universe, where AGN activity is generally suppressed in the densest regions, such as galaxy clusters, due to environmental quenching and gas stripping \citep[e.g.,][]{Kauffmann2004}. The enhanced AGN fraction at high redshift has been interpreted as evidence that black hole fueling is more efficient at earlier cosmic times, likely driven by processes such as mergers and interactions \citep[e.g.,][]{Lemaux2022} but at least in our current sample we are not able to confirm this with ODIN.

\section{Results} \label{sec:results}

\subsection{Dependency with Ly$\alpha$ luminosity and UV magnitude}

To investigate the relationship between LAE multiplicity and intrinsic galaxy properties, we analyze the mean Ly$\alpha$ luminosity and rest-frame UV magnitude of LAEs within identified multiples.

\begin{figure*}
\centering
\includegraphics[width=0.8\textwidth]{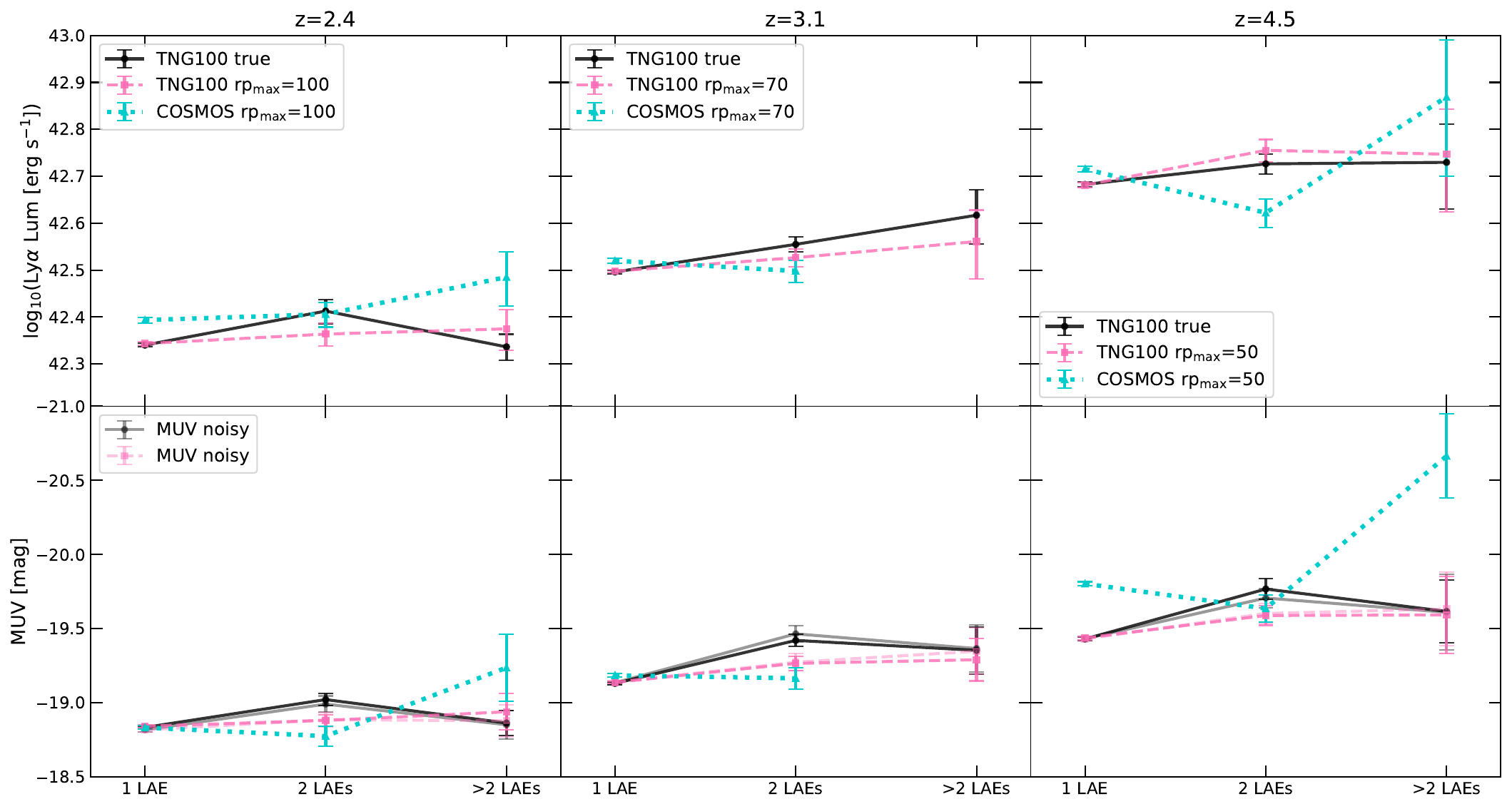}
\caption{Mean Ly$\alpha$ luminosity (\textit{top panels}) and UV magnitude (\textit{bottom panels}) as a function of the LAE multiplicity, for redshifts $z = 2.4$ (\textit{left}), $z = 3.1$ (\textit{middle}) and $z=4.5$ (\textit{right}). Solid black lines correspond to the TNG100-true multiples, dashed pink lines show the TNG-identified multiples (with $rp_{\text{max}} = 100$ pkpc at $z = 2.4$, 70 pkpc at $z = 3.1$, and 50 pkpc at $z = 4.5$), and dotted cyan lines correspond to the ODIN-COSMOS sample under the same selection criteria. The lighter black and pink lines indicate the mean UV magnitudes perturbed by photometric uncertainties for the mock samples. Error bars represent the standard error of the mean. % error_mean = std/sqrt(cantidad)
Overall, both Ly$\alpha$ luminosity and UV magnitude tend to increase slightly with LAE multiplicity, supporting the idea that multiple LAEs trace more massive and actively star-forming halos. 
}
\label{fig:lya_lum_all}
\end{figure*}

The top panels of Figure \ref{fig:lya_lum_all} show the mean Ly$\alpha$ luminosity of individual LAEs as a function of the number of LAEs in multiples at the three redshifts: $z = 2.4$ (left), $z = 3.1$ (middle), and $z = 4.5$ (right). 
We compare the results for the TNG100-true (solid black lines), TNG100-identified (dashed pink lines), and ODIN-COSMOS (dotted cyan lines) datasets, using the previously defined $rp_{max}$. To ensure a fair comparison, we consider only the ODIN-COSMOS multiples with Ly$\alpha$ luminosities below 10$^{43.5}$ erg s$^{-1}$, thus avoiding possible AGN contaminants.

We find that the mean Ly$\alpha$ luminosity either shows a slight tendency to increase with multiplicity or remains roughly constant in the mock samples (except for the TNG100 true sample at $z = 2.4$, where it decreases for multiples with more than two LAEs). In the observational data, this trend is more evident at $z = 2.4$ and $z=4.5$, suggesting that higher multiplicity systems at these redshifts are preferentially associated with more luminous LAEs, assuming the incompleteness and contamination are similar to those in the mock.
We also performed a comparison with a random sample of pairs and triples drawn from the isolated LAEs, finding that their mean luminosities and magnitudes remain constant. This supports the interpretation that the observed variations in the real sample reflect a genuine trend. However, it should be noted that the dispersion introduced by the error bars prevents us from establishing a conclusive trend, particularly for the TNG100-true sample at $z = 2.4$ which shows an opposite behavior.

The offset between the mock and ODIN-COSMOS mean values could reflect the effects of contamination and incompleteness in the identification process, as discussed earlier.

Both the halo mass (Fig. \ref{fig:masas_nlae}) and the mean Ly$\alpha$ luminosity show an increase with redshift. This result, however, seems to be at odds with \citet{Khostovan2019} and \citet{Herrero-Alonso2023} who find no significant redshift evolution in the typical halo masses hosting LAEs, indicating that LAEs reside in halos of similar mass across cosmic time. Moreover, \citet{Herrero-Alonso2023} points to the possibility that the observed clustering dependence on Ly$\alpha$ luminosity is driven by cosmic evolution of the clustering of objects of similar mass. 
%Still, our results are further supported by the observation that the trend of Ly$\alpha$ luminosity with LAE multiplicity does not show a clear dependence on redshift, a trend that is more robustly present in the simulation than in the ODIN data. 

The bottom panels of Fig. \ref{fig:lya_lum_all} display the mean M$_\text{UV}$ magnitudes as a function of LAE multiplicity. A similar trend is observed in the mock samples: LAEs show a weak tendency to be more UV-luminous as the multiplicity increases.
This indicates that more massive halos hosting higher LAE multiplicities could also drive more vigorous star formation.
However, the ODIN-COSMOS sample shows a significant dependence on the LAEs of multiples with more than 2 LAEs.
We also show, in gray and lighter pink, the mean noisyfied M$_\text{UV}$ magnitudes (as described in Section~\ref{sec:lae_samples}) for the TNG100-true and TNG100-identified samples, and they follow similar trends as the noise-free magnitudes.

%There is a discrepancy between the ODIN-COSMOS data and TNG100 predictions in M$_\text{UV}$, which shows an offset of $\sim 0.5$ magnitude difference, or a $60\%$ luminosity excess in the mock. Although the \cite{Vogelsberger2020} model does not reproduce the UV magnitudes at 1600 \AA, it is interesting that the \cite{DijkstraWyithe2012} model is still able to reproduce the Ly$\alpha$ line luminosities without an offset. 

%Additionally, the observed M$_\text{UV}$ values are subject to significant uncertainties, as they are indirectly derived from Ly$\alpha$ fluxes and REWs, both of which carry their own measurement errors. These propagate into the UV continuum estimates and may partially account for the aparent offset with the simulation and a less clear trend with multiplicity. On the simulation side, the UV magnitudes are computed based on stellar population synthesis models and dust prescriptions \citep{Vogelsberger2020}, which may oversimplify the complexity of radiative transfer and attenuation in real galaxies. This could also contribute to the mismatch between observed and simulated UV properties, particularly at the faint end.

%Our samples of multiple LAEs show trends that are coincident between mock and observations in terms of the increase of mean Ly$\alpha$ and UV luminosity as a function of number of LAEs per multiple. 
Our samples of multiple LAEs appear to exhibit broadly consistent trends between the mock and observational data, with a possible increase in the mean Ly$\alpha$ and UV luminosities as the number of LAEs per multiple rises. These multiples represent physical associations of star forming galaxies at relatively high redshifts, where star formation is at its densest in the history of the universe. We will therefore attempt to characterize the local conditions where these LAEs reside in the following subsection, and in the subsequent one, we propose a method to test the physical connection between LAE multiples and star formation due to the presence of perturbing neighbors.

\subsection{Group-wide Ly$\alpha$ and UV luminosity surface densities}

In this section, we analyze the virial projected Ly$\alpha$ and UV luminosity densities of LAEs, defined as the sum of the luminosities of all galaxies in a multiple, divided by an estimate of the projected area of the halo.  These quantities can help to gain insights into the star formation processes within virialized halos and their relation to the presence of physically associated neighbors. By examining the correlation between the local UV luminosity virial density, and observable quantities, such as the number of LAEs or the relative fraction of LAEs among halo galaxies (accessible in simulations), we hope to obtain insights into how star formation activity is distributed within halos and whether the presence of LAEs in multiples affects this distribution.

\begin{figure*}
\centering
\includegraphics[width=0.8\textwidth]{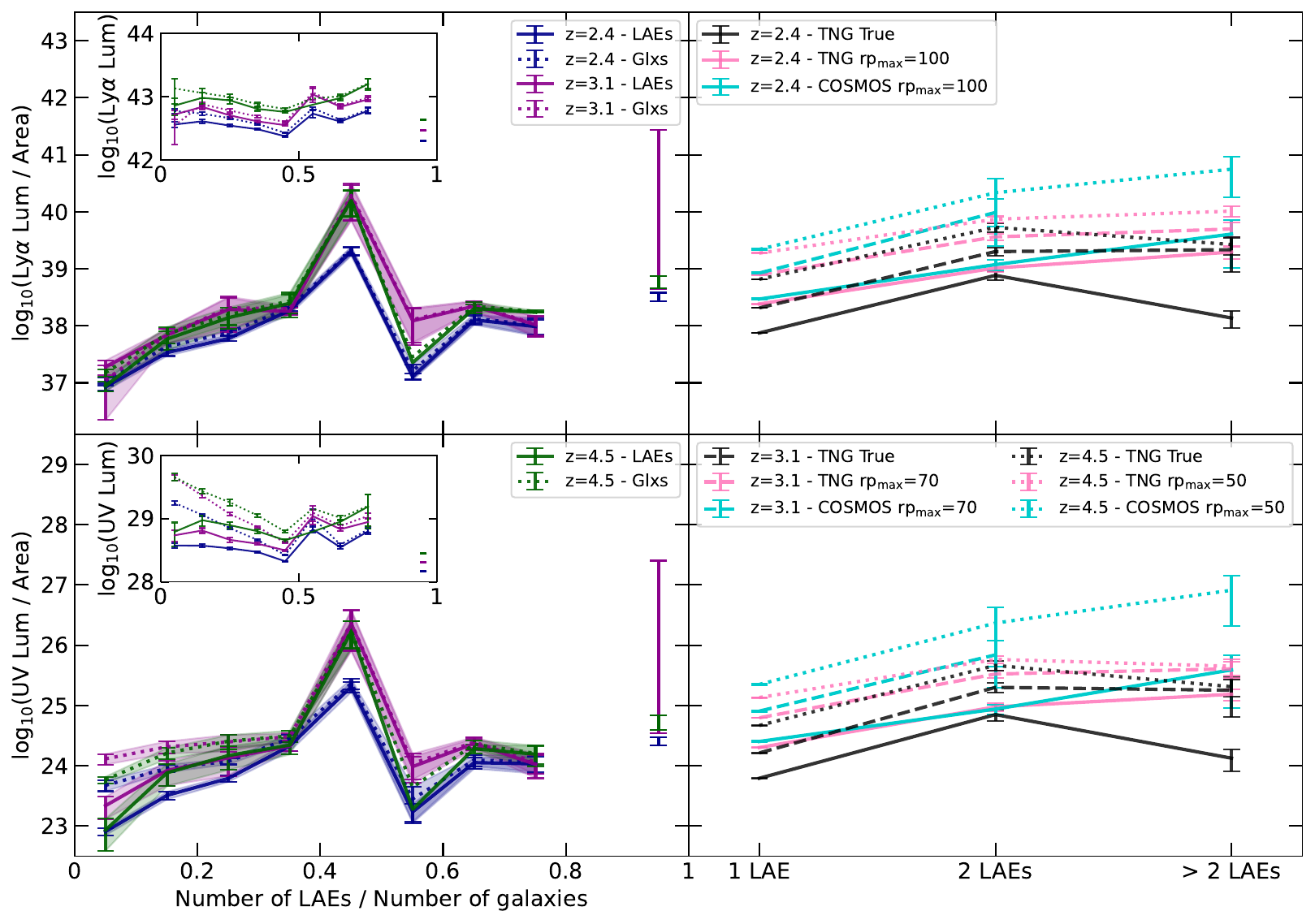}
\caption{Virial surface brightness of Ly$\alpha$ (erg s$^{-1}$ kpc$^{-2}$) and UV (erg s$^{-1}$ Hz$^{-1}$ kpc$^{-2}$) luminosities as a function of LAE multiplicity.
\textit{Left panels:} Mean virial surface luminosities for Ly$\alpha$ (\textit{top}) and UV (\textit{bottom}) emissions in TNG100, defined as the total luminosity per projected area, at $z = 2.4$ (blue), $z = 3.1$ (magenta) and $z=4.5$ (green), plotted as a function of the fraction of LAEs relative to the total number of galaxies in the halo. Solid lines correspond to LAE-based densities, while dotted lines correspond to galaxy-based densities. Insets display the mean total luminosity before normalization.
Error bars represent the standard error of the mean. Both Ly$\alpha$ and UV virial densities increase with the LAE fraction, reaching a peak at $\sim0.5$. This behavior is mainly driven by the Ly$\alpha$ model, which produces a more compact spatial distribution of LAE galaxies within halos.
\textit{Right panels:} Mean virial surface luminosities as a function of LAE multiplicity. Black lines show the TNG100-true values, pink lines the TNG100-identified sample, and cyan lines the ODIN-COSMOS data, with solid, dashed, and dotted lines denoting $z = 2.4$, $z = 3.1$, and $z=4.5$, respectively.
Error bars represent the standard error of the mean in each bin.
Higher LAE multiplicity appears to be associated with increased virial surface brightness densities.}
\label{fig:lum_uv_a_ngal}
\end{figure*}

We define the virial luminosity density, corresponding to the total Ly$\alpha$ (or UV) luminosity of LAEs or galaxies, as the sum of their individual luminosities divided by the projected area $\pi \cdot (\text{dist}_{max}/2)^2$, computed using two distinct but closely related spatial scales in the simulations, and the one available to ODIN for observations:

- The \textit{physical-scale} corresponds to the maximum 3-dimensional radius that contains all galaxies within a given halo. We define $\text{dist}_{max}$ as the largest 3D separation between any two halo member galaxies. This scale provides a comprehensive view of the luminosity distributions across the full extent of the halos in the mock and are presented in the left panels of Figure \ref{fig:lum_uv_a_ngal}. 

- The \textit{observational-scale} is defined observationally (in the mock and in ODIN), where $\text{dist}_{max}$ represents the maximum projected radius that encloses the galaxies identified as part of a LAE multiple. This approach uses observationally available information and aligns directly with ODIN data, where only LAEs are mapped, ensuring consistency in system-size measurements.
For high multiplicities, $\text{dist}_{max}$ is defined as the maximum projected separation between all LAEs, corrected by a factor in the case of pairs to account for central–satellite and satellite–satellite configurations which we obtain from the mock catalog. For isolated LAEs, $\text{dist}_{max}$ is set as the mean projected separation of the two-LAE multiples.  
This observational-scale analysis is shown in the right panels of Fig. \ref{fig:lum_uv_a_ngal}.

The left panels of Fig. \ref{fig:lum_uv_a_ngal} show the mean total Ly$\alpha$ luminosity (top panel) and UV luminosity (bottom) of normal galaxies (dotted lines) and LAEs (solid lines) in the mock catalog at $z=2.4$ (blue), $z=3.1$ (magenta) and $z=4.5$ (green), normalized by the physical-scale area, as a function of the number of LAEs relative to the total number of galaxies in the halo. 
At high LAE fractions ($> 0.4$), the densities traced by all galaxies closely follow that traced by LAEs, indicating that halos with high LAE fractions are predominantly UV- or Ly$\alpha$-luminous due to LAEs themselves. At lower fractions, a divergence emerges, especially in the case of UV, reflecting contributions from non-LAE galaxies. The agreement is tighter in Ly$\alpha$ across the entire LAE fraction range. 

For LAEs, both Ly$\alpha$ and UV densities increase with increasing LAE fraction, peaking at around $\sim 0.5$, although errorbars for this fraction are large as this point receives contributions from halos with pairs or a few LAEs. Errors diminish for larger fractions which correspond to halos with larger numbers of LAEs. 
This behavior suggests that the adopted Ly$\alpha$ luminosity model predicts a higher fraction of LAEs in regions with enhanced UV radiation density from star formation — at least up to halos with LAEs making up half the halo population. 
Notably, this trend is not driven by changes in the average UV or Ly$\alpha$ luminosity of LAEs and normal galaxies. As shown in the insets, the mean total Ly$\alpha$ luminosity remains approximately constant, while the UV luminosity slightly decreases with increasing LAE fraction.
This highlights that the increase in surface densities arises primarily from the Ly$\alpha$ model tending to produce a more compact spatial distribution of LAE galaxies, rather than from intrinsic luminosity enhancements.

For our previously defined sample of LAE multiples, the right-hand panels of Fig. \ref{fig:lum_uv_a_ngal} show the mean total Ly$\alpha$ luminosity (top) and UV luminosity (bottom) of LAEs within each multiple, normalized by the observational-scale area, as a function of LAE multiplicity at $z = 2.4$ (solid lines), $z = 3.1$ (dashed lines) and $z = 4.5$ (dotted lines).
In both cases, the virial surface density values are systematically higher at $z=4.5$ than at $z = 3.1$, which in turn are higher than at $z = 2.4$, suggesting more intense star formation activity or more compact associations at earlier cosmic times for both ODIN-COSMOS and TNG100 multiples.

Both the TNG100-identified sample and the ODIN-COSMOS data also exhibit a positive trend, with higher LAE multiplicity associated with increased virial surface brightness densities. This supports the hypothesis that environments hosting multiple LAEs are more actively star-forming and/or denser, which enhances both Ly$\alpha$ and UV emission densities.
However, the TNG100-true sample shows a different behavior, particularly at $z = 2.4$. In this case, the projected separations ($rp_{max}$) of true systems can reach up to $\sim700$ pkpc. Due to our imposed $rp_{max}$ cut for multiples identification, many of these extended multiples are split into multiple smaller multiples or identified as isolated LAEs, underestimating the true multiplicity. As a result, the surface brightness appears to decline at high multiplicities. At $z = 3.1$ and $4.5$, halos are generally more compact, and this issue is mitigated, leading to a more consistent trend with increasing multiplicity. Still, the increase in luminosity density of identified multiples could be an artifact of the identification rather than a physical trend. 

It is important to note that this trend could be affected by the method adopted to define the radius of the virialized object.  However, we find that it is also present when the normalization is defined using the average value of the halo radius, R$_{200}$ (see Fig. \ref{fig:virial_densities} in Appendix \ref{appen:virial_densities}) corresponding to each sample.  

It should still be noticed the excellent consistency between the mock and ODIN observations, which supports the validity of the Ly$\alpha$ emission model by \cite{DijkstraWyithe2012} in accurately reproducing the observed properties of LAE multiples within dark matter halos (with our samples). This agreement suggests that the model effectively captures the key physical processes governing Ly$\alpha$ radiative transfer and star formation activity in dense environments, such as those traced by multiple LAEs in the same halo.

With larger samples such as the full ODIN survey, it may be possible to have a better sample of multiples with many LAEs. Confirming an increasing or decreasing trend of virial UV and Ly$\alpha$ luminosity density in multiples, unaffected by the low number statistics and the fragmentation of halos into pairs or triplets, could help put star formation in the framework of disk instabilities set off by perturbers.

The insights of \cite{MoMaoWhite1998}, which point that bursts of star formation can take place in flattened gas disks because they are susceptible to perturbations that can ignite star formation and consume the gas within them, could be put to the test with the full ODIN survey, given the expected high star formation activity of LAEs. In a halo containing multiple galaxies (subhalos), it is plausible that interactions among these galaxies act as the triggers for such perturbations, leading to significant hydrogen line emissions, as is characteristic of Ly$\alpha$ emitting galaxies. Alternatively, the triggering of star formation could also be due to direct infall of cold gas, as proposed by \cite{Dekel}.  We next make an attempt to tell these two possible effects apart with the mock catalog.

\subsection{Star formation due to instabilities triggered by members of LAE multiples} \label{sec:model}

We test this idea using the mock LAE samples extracted from TNG100, an analysis that is further motivated by the apparent excess of disk morphologies seen in galaxies at the ODIN-COSMOS redshifts as revealed from optical rest-frame imaging by JWST (e.g. \citealt{Kuhn_2024}). It is important to note that the degree of star formation enhancement resulting from such perturbations may vary across different simulations, depending on the specific numerical implementations of the star formation models. While perturbations can generally induce local density peaks that may promote star formation, the extent of this enhancement is sensitive to the adopted subgrid physics and feedback prescriptions.

We start assuming that all galaxies within the halo experience a perturbation capable of triggering a star-forming event when the perturber induces a change in the disk gravitational potential proportional to the disk's own gravitational potential by a factor $f_\phi$, 
\begin{equation}
\Delta \phi \sim f_\phi \phi_d = f_\phi \frac{G M_d}{R_d},
\label{eq:disk}
\end{equation}
where $\Delta \phi$ and $\phi _d$ are the perturbation and disk gravitational potentials, respectively, and we write the disk potential as a function of its mass ($M_d$) and radius ($R_d$).

Assuming that subhalos have a minimum mass ($M_{sh,min}$) and taking advantage of the steep slope of the subhalo mass function, we adopt the approximation that only subhalos with this minimum mass are present in the halo. Under this assumption, the perturbation can be approximated by the gravitational potential of these subhalos,
\begin{equation}
\Delta \phi \sim \frac{G M_{sh,min}}{\langle d_{sh,min} \rangle},
\label{eq:sh}
\end{equation}
where $\langle d_{sh,min} \rangle$ is the mean distance between substructures.

Combining Eqs. \ref{eq:disk} and \ref{eq:sh}, we find,
\begin{equation}
M_{sh,min} \sim f_\phi \frac{M_d}{R_d} \langle d_{sh,min} \rangle.
\label{eq:Msh}
\end{equation}

Considering $V_{sh,min}$ a cubic volume with a side length of $\langle d_{sh,min} \rangle$, and the mass (or volume) of the halo can be approximated as the product of the number of subhalos and their respective mass (or volume), we can express,
\begin{equation}
M_{sh,min} \sim \left(f_\phi \frac{M_d}{R_d} (200 \rho_c(z))^{-1/3} \right)^{3/2}
\label{eq:Msh2}
\end{equation}
where $\rho_c(z)$ is the critical density at that redshift.

For our calculations, we assume that the mass and radius of the disk are proportional to the mass and radius of the subhalo, following the relations $M_d = f_M  M_{sh}$ and $R_d = f_R R_{sh}$, where $M_{sh}$ is the total mass for the subhalo and $R_{sh}$ is the subhalo half mass radius. Consequently, Equation \ref{eq:Msh2} depends on the perturbation factor $f_\phi$ and the ratio of the proportionality factors $f_M / f_R$.
This ratio $f_M / f_R$ reflects the relationship between the fraction of the subhalo mass contributing to the galaxy's disk ($f_M$) and the fraction of the subhalo's radius defining the galaxy's disk size ($f_R$). This parameter essentially encapsulates how the properties of the subhalo translate into the disk characteristics, such as its gravitational potential and physical extent. This approach allows us to explore the relationship between subhalo interactions and, in our particular case, LAEs.

Using the values of $M_{sh}$ and $R_{sh}$ for the galaxies from TNG100, together with $M_d$ and $R_d$ from the literature, we can obtain a simple estimate of the ratio $f_M / f_R$. For this, we adopt the results of \cite{Ma2024}, who measured the cold gas mass and half-mass radius for different galaxy samples at different redshifts in TNG50. For consistency with our analysis, we use their most massive sample (referred to in their paper as M4), which matches the properties of the LAEs in TNG100.
%While their study relied on TNG50 instead of TNG100, we consider that using their most massive subsample minimizes the impact on our selection.
We obtain the mean values of $f_M / f_R =$ 2.43, 0.78, and 0.63 at reshifts 2.4, 3.1, and 4.5, respectively. Using these values and varying the perturbation factor $f_\phi$, we can test how well our model reproduces the distribution of subhalo masses and gain insight into the role of subhalos in driving galaxy evolution.

\begin{figure}
\centering
\includegraphics[width=0.49\textwidth]{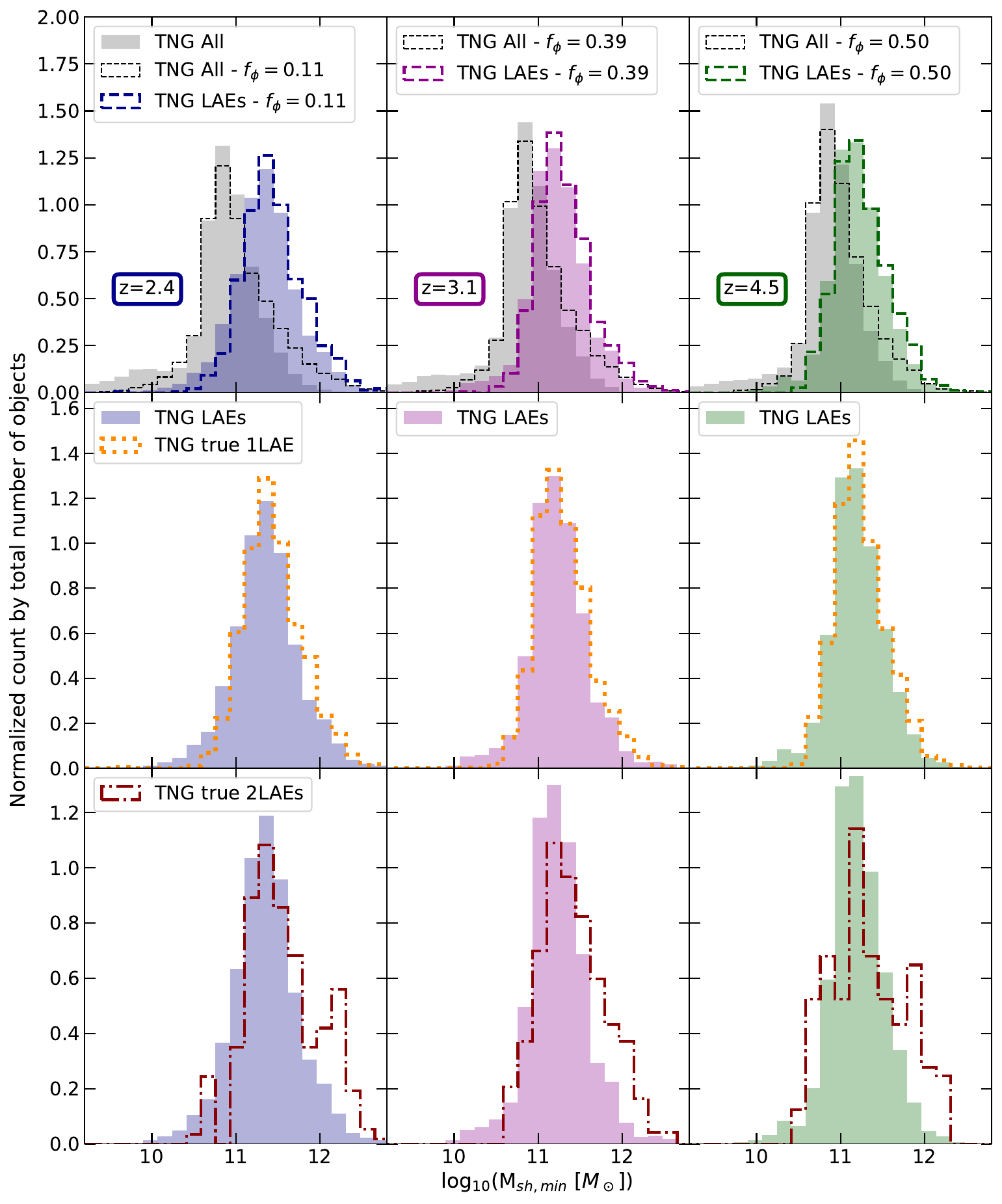} 
\caption{Distribution of minimum subhalos masses in the halos at $z=2.4$ (\textit{left panel}), $z=3.1$ (\textit{middle panel}) and $4.5$ (\textit{right panel}). 
In the \textit{top panels}, the minimum subhalo mass distributions from TNG100 are shown as shaded black histograms, while the model predictions described in Section \ref{sec:model} are overplotted as black dashed lines (with $f_\phi = 0.11$ and $f_M / f_r =2.43$ at $z=2.4$, $f_\phi = 0.39$ and $f_M / f_r =0.78$ at $z=3.1$, and $f_\phi = 0.50$ and $f_M / f_r =0.63$ at $z=4.5$). 
Using these factors, we test the model against all TNG100 LAEs (colored dashed lines in the \textit{top panels}), TNG100 true 1 LAEs (orange dotted lines in the \textit{middle panels}), and TNG100 true 2 LAEs (red dash-dotted lines in the \textit{bottom panels}), comparing them with the corresponding minimum subhalo mass distributions of TNG100 LAES (shaded colored histograms).
The model shows a good agreement with the minimum subhalo mass distribution for all galaxies, for all LAEs, and isolated LAEs, but struggles to reproduce as closely the distribution for LAE pairs.}
\label{fig:sh_model}
\end{figure}

Figure \ref{fig:sh_model} presents the resulting distributions of the minimum subhalo masses obtained using our model with TNG100 and its various subsets, at $z=2.4$ (left panels), $z=3.1$ (middle panels), and $z=4.5$ (right panels). For comparison, the top panels display the distribution of minimum subhalo masses for all TNG100 halos as shaded black histograms. 
The model predictions for the full galaxy sample are shown as black dashed lines, using the $f_M / f_r$ values mentioned above and the $f_\phi$ values that minimize the difference between the expected and model histograms. We find $f_\phi = 0.11$, $0.39$, and $0.50$ at $z=2.4$, $3.1$, and $4.5$, respectively. 
The model generally shows good agreement with the actual masses of subhaloes in the simulation, but tends to underpredict the low-mass tail, which is largely introduced by satellite galaxies that the model fails to reproduce. In addition, it overpredicts the number of high-mass central subhaloes, leading to an artificial excess at the high-mass end. 
The perturbation factor $f_\phi$ increases with redshift, indicating that stronger perturbations are required to trigger star formation at earlier cosmic times.  Note that it is quite unlikely that a perturber will induce $f_\phi>0.1$ as this would imply strong interactions. We interpret the values for this quantity obtained for $z=3.1$ and $4.5$ as indications that, at least in the simulation, the instabilities driving star formation are not due to interactions with other satellites (in line with \citealt{Dekel}).
Colored shaded histograms show the minimum subhalo masses for halos hosting LAEs, while the colored dashed lines represent the model predictions for all TNG100 LAEs using the same scaling factors. In this case, the fit improves significantly, as LAEs do not exhibit the low- or high-mass tails seen in the full halo sample.
In the middle and bottom panels, we compare the minimum masses of the LAE subhalos with the model applied to the TNG100 True multiples with 1 (orange dotted, middle panels) and 2 (red dash-dotted, bottom panels) LAEs, with the same parameters.

It is significant that the model successfully reproduces the distribution of the minimum subhalo mass present in the simulation at $z=2.4$ with a reasonable level required for the perturbation of the potential. This alignment suggests that indeed the simulation effectively traces subhalo-driven perturbations at $z\lesssim3$, which are critical for triggering star formation. Such consistency is expected, as the star formation prescription in IllustrisTNG includes the requirement that the gas forming stars must be in a locally collapsing state. This condition naturally aligns with the onset of instabilities, which can be induced by intra halo perturbations among other processes such as cold gas inflows. Therefore, the ability to match $M_{sh,min}$ in the mock at these lower redshifts lends confidence to both the simulation's physical modeling and our interpretation of a shift from subhalo driven to cold flows perturbations at $z>3$ a crucial test that will hopefully be achievable with the full ODIN survey.  

It is interesting to note that the inferred $f_\phi$ perturbation factor can be used to fit the underlying minimum subhalo mass distribution for all galaxies, LAEs, and LAEs in single multiples. The agreement between single LAEs -- not necessarily isolated as they can have non-LAE satellites -- and the overall LAE distribution is largely reproduced, as isolated LAEs constitute the majority of the sample. In contrast, applying the same model values to pairs of LAEs yields multimodal distribution but still in reasonable agreement, with the slight discrepancies showing signs of an increase at higher redshifts. This result again suggests that the presence of a neighboring LAE influences star formation activity at $z=2.4$, at least within the framework of this model.

The minimum mass threshold which is reproduced by our model with a reasonable level of perturbation of the potential for $z=2.4$ also hinges on the fraction of LAEs that we select in order to reproduce the abundance of observed LAEs. In future work one can improve the model also taking into account the expected recurrence timescale of such events. In halos hosting multiple subhalos above $M_{sh,min}$, the typical orbital periods of these companions could naturally lead to quasi-periodic episodes of enhanced star formation and Ly$\alpha$ emission that would explain the fraction that we apply by hand. In this framework, the observed multiplicity of LAEs and their visible fraction could be seen as a tracer of  the characteristic duty cycle of their Ly$\alpha$ -bright phases.

\section{Conclusions} \label{sec:conclusions}

In this study, we used data from the ODIN survey—the largest narrowband filter survey conducted to date—which employs three custom-designed narrowband filters to detect LAEs across three equally spaced epochs in cosmic history. Using the LAE samples at redshifts 2.4, 3.1, and 4.5, we searched for multiple LAE associations as a proxy for halo masses. This approach was supported by an analysis using a model \citep{DijkstraWyithe2012} that reproduces the LAE population, applied to the IllustrisTNG100 simulations \citep{Andrews2025}. These simulations allowed us to test our selection method and test for systematic effects present in the comparison. To this end, we constructed a sample of LAEs that is consistent between simulations and ODIN observations. Our main results are listed below.

\begin{enumerate}
    \item In the mock catalog, we find that it is possible to find physical multiples of LAEs, with controlled contaminants and reasonable completeness. The resulting multiples show a clear correlation between the number of LAEs and the mass of their host dark matter halos. Although most LAEs are isolated, systems with higher multiplicity tend to reside in more massive halos. Our method for selecting LAE multiples successfully recovers this trend to a significant degree.
    \item In both ODIN and mock, we find indications that the mean Ly$\alpha$ luminosity and UV magnitudes of LAEs in the multiples increase with multiplicity, supporting a connection between Ly$\alpha$ emission, multiplicity, and halo mass. However, this trend is not conclusive given the dispersion of the error bars, and further statistical analysis would be required to confirm or refute it. %Trends in UV magnitudes with multiplicity are less pronounced and exhibit greater discrepancies between the observational and simulated samples.
    \item  UV and Ly$\alpha$ virial (or halo-wide) surface brightness densities increase with LAE multiplicity, suggesting that more massive halos trace denser, more actively star-forming regions. Tests with the mock catalog suggest that this trend seems to be  driven by a spatial compactness of LAEs and not by higher individual luminosities. We also find that LAEs are able to trace the UV virial density in an unbiased way. A redshift dependence is observed, with higher halo-wide surface brightness densities at $z=4.5$ and $z = 3.1$ than at $z = 2.4$, consistent with more vigorously star-forming environments at earlier cosmic times.
    \item We find a strong agreement between the modeled and ODIN-COSMOS fractions of singles and multiples with 2 and with more members, as well as consistent Ly$\alpha$ properties. This validates the Ly$\alpha$ emission model by \citet{DijkstraWyithe2012} and \citet{Andrews2025}, confirming its ability to reproduce the spatial and radiative characteristics of LAEs within dark matter halos. %However, the model produces slightly lower UV magnitude ($\sim0.5mag$) at 1600 \AA, particularly at $z<3.1$.
    \item Our subhalo-based model for perturbation triggered star formation events accurately reproduces the minimum subhalo mass distribution of TNG100 LAEs for the $z=2.4$ sample, supporting the idea that intra-halo perturbations —rather than LAE companions— are key to triggering star formation in these galaxies.  Interestingly, at higher redshifts we find that the required level of perturbation suggests that in TNG100, star formation is likely triggered by processes other than gravitational perturbations by subhalos.
\end{enumerate}

Ultimately, our findings from the ODIN survey, corroborated by IllustrisTNG100 simulations, robustly establish Ly$\alpha$ emitter multiplicity as a powerful and observationally accessible proxy for tracing dark matter halo masses and the underlying processes of star formation in the distant universe, thereby significantly advancing our understanding of galaxy evolution in cosmic dawn environments.

The full ODIN survey will not only provide a larger sample of LAEs, but also a wide set of physical properties derived from SED fitting, such as stellar mass and age, as well as structural parameters like galaxy size and morphology from HST or JWST imaging. The availability of these ODIN LAE data will enable future analyses of correlations with additional physical properties.

\begin{acknowledgements}
This work utilizes observations at Cerro Tololo InterAmerican Observatory, NSF’s NOIRLab (Prop. ID 2020B0201; PI: K.S.L.), which is managed by the Association of Universities for Research in Astronomy under a cooperative agreement with the National Science Foundation.
This work has been partially supported with grants from Agencia Nacional de Promoci\'on Cient\'ifica y Tecnológica, the Consejo Nacional de Investigaciones Cient\'{\i}ficas y T\'ecnicas (CONICET, Argentina) and the Secretar\'{\i}a de Ciencia y Tecnolog\'{\i}a de la Universidad Nacional de C\'ordoba (SeCyT-UNC, Argentina).
MCC acknowledges support from a CONICET Doctoral Fellowship. NP acknowledges support from PICT 2021-0700 and PICT Raices Federal 2023-0002.
This material is based upon work supported by the NSF Graduate Research Fellowship Program under Grant No. DGE-2233066 to NF. EG and NF acknowledges support from NSF grant AST-2206222 and NASA Astrophysics Data Analysis Program grant 80NSSC22K0487.
KSL acknowledges financial support from the National Science Foundation under grant Nos. AST-2206705 and AST-2408359, and from the Ross-Lynn Purdue Research Foundations.
CG acknowledges support from NSF AST-240538.
LG also gratefully acknowledges financial support from ANID - MILENIO - NCN2024$\_$112, from FONDECYT regular project number 1230591, and from the ANID BASAL project FB210003.
HSH acknowledges the support of Samsung Electronic Co., Ltd. (Project Number IO220811-01945-01), the National Research Foundation of Korea (NRF) grant funded by the Korea government (MSIT), NRF-2021R1A2C1094577, and Hyunsong Educational \& Cultural Foundation.
SL acknowledges support from the National Research Foundation of Korea(NRF) grant (RS-2025-00573214) funded by the Korea government (MSIT).
\end{acknowledgements}

% WARNING
%-------------------------------------------------------------------
% Please note that we have included the references to the file aa.dem in
% order to compile it, but we ask you to:
%
% - use BibTeX with the regular commands:
   \bibliographystyle{aa} % style aa.bst
   \bibliography{bibliography} % your references Yourfile.bib
%
% - join the .bib files when you upload your source files
%-------------------------------------------------------------------

% \begin{thebibliography}{}

%   \bibitem[Baker(1966)]{baker} Baker, N. 1966,
%       in Stellar Evolution,
%       ed.\ R. F. Stein,\& A. G. W. Cameron
%       (Plenum, New York) 333

%    \bibitem[Balluch(1988)]{balluch} Balluch, M. 1988,
%       A\&A, 200, 58

%    \bibitem[Cox(1980)]{cox} Cox, J. P. 1980,
%       Theory of Stellar Pulsation
%       (Princeton University Press, Princeton) 165

%    \bibitem[Cox(1969)]{cox69} Cox, A. N.,\& Stewart, J. N. 1969,
%       Academia Nauk, Scientific Information 15, 1

% \end{thebibliography}

\begin{appendix}

\section{Contamination and completeness of LAE multiples} \label{appen:test_cyc}

This Appendix presents the definitions and methodologies used to quantify contamination and completeness, which are applied in Section \ref{sec:optimization} to test our selection method for LAE multiples.

\begin{figure}[!ht]
\centering
\includegraphics[width=0.49\textwidth]{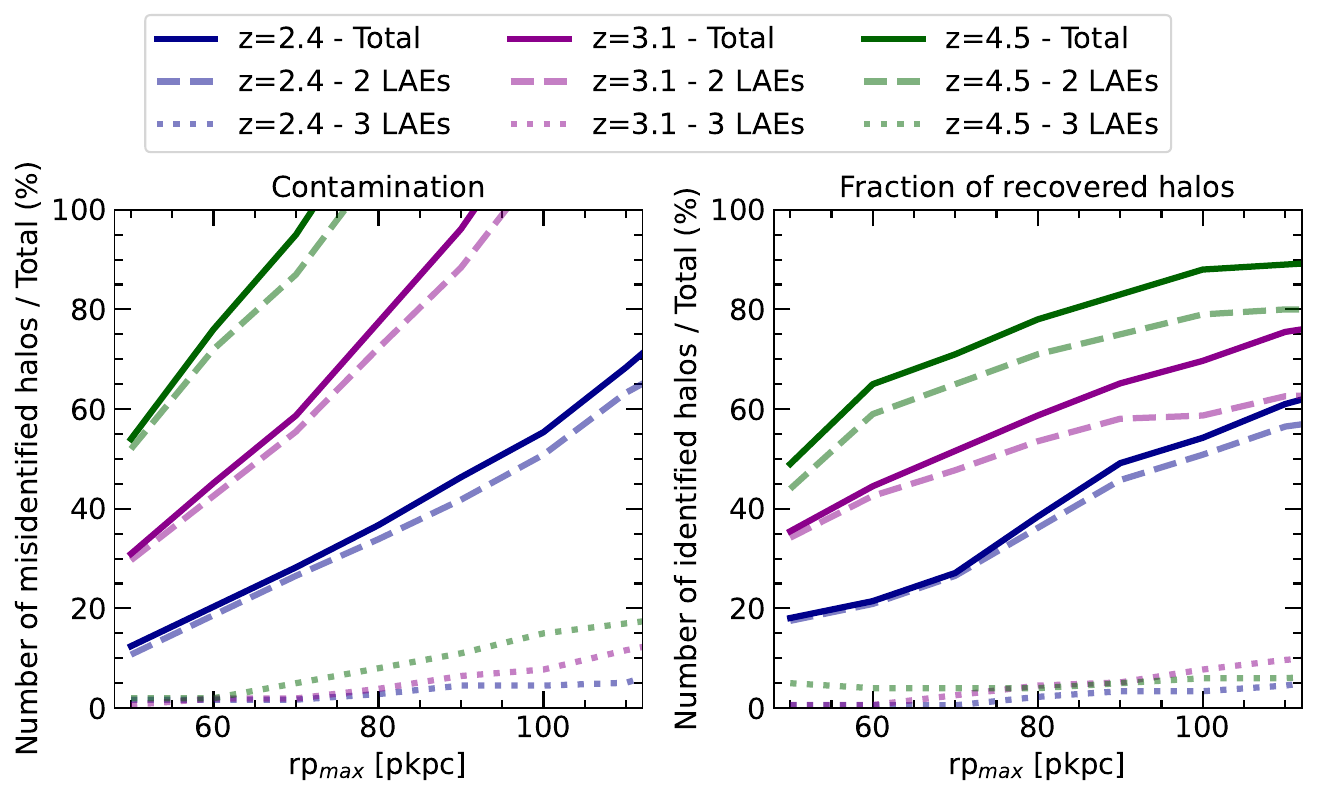} 
\caption{Fraction of detections in which not all members belong to the same halo (contamination, \textit{left panel}) and fraction of halos with at least two galaxies that are recovered (completeness, \textit{right panel}), as a function of the maximum projected distance $rp_{max}$, for the mock samples at $z = 2.4$ (blue), $z = 3.1$ (magenta) and $z=4.5$ (green). The total fractions are shown with solid lines, while the fractions corresponding multiples of 2 LAEs are shown with dashed lines, and multiples of 3 LAEs with dotted lines.
Both the contamination and the fraction of recovered halos increases with the $rp_{max}$, so it is necessary to find a compromise value between them.}
\label{fig:cyc}
\end{figure}

We define contamination as the fraction of detections in which not all members belong to the same halo (i.e., incorrectly identified associations), relative to the total number of true multiples. The left panel of Figure \ref{fig:cyc} shows the contamination at the three redshifts ($z=2.4$ in blue, $z=3.1$ in magenta, and $4.5$ in green). The solid lines represent all recovered multiples, while the dashed and dotted lines show the fractions corresponding to associations with 2 and 3 LAEs, respectively. As associations with 2 LAEs constitute the majority, the dashed line closely follows the total.
We observe that contamination grows dramatically with projected distance, as expected due to the depth of the data, with only the smallest projections ensuring that the contamination is not unreasonably high. This analysis also reveals that multiples with more LAEs exhibit lower contamination, indicating they are more reliable tracers and more likely to reside within the same halos. However, such multiples are relatively rare in the present ODIN sample we analyze here.

On the other hand, we define the completeness as the fraction of recovered halos (with at least two galaxies). This is shown in the right panel of Figure \ref{fig:cyc} as a function of the maximum projected distance (solid lines). As expected, this fraction increases with projected distance.

\section{Analysis of LAEs (central and satellite) in IllustrisTNG} \label{appen:frac_of_laes}

LAEs may not fully capture the diversity of high-redshift galaxy populations. In TNG100, only 15$\%$, 17$\%$, and 18$\%$ of galaxies are classified as LAEs at $z = 2.4$, $z = 3.1$, and $z=4.5$, respectively, according to our LAE criteria (i.e., Ly$\alpha$ luminosity above the adopted threshold and REW $\geq$ 20 \r{A}).
Here, we analyze the fractions of LAEs relative to the overall galaxy population within halos and investigate how these fractions vary with different halo and galaxy properties.

\begin{figure}[]
\centering
\begin{subfigure}
\centering
\includegraphics[width=0.49\textwidth]{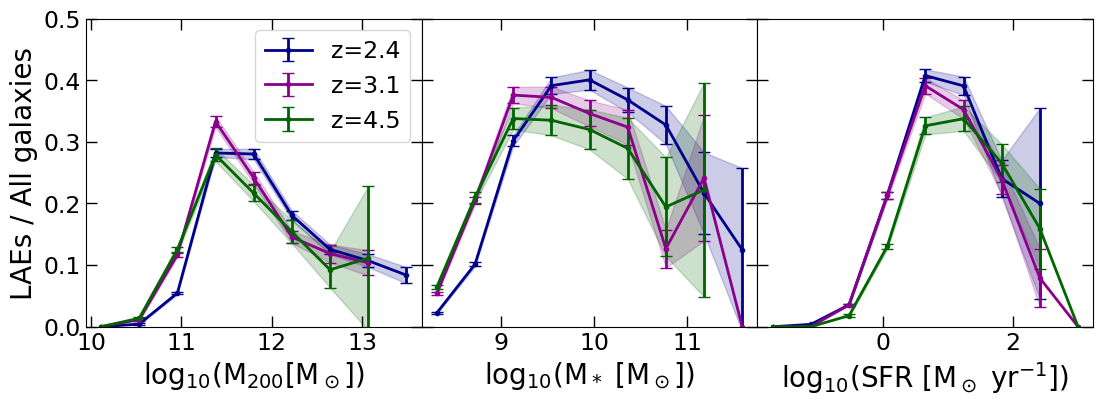}
\end{subfigure}
\hfill
\begin{subfigure}
\centering
\includegraphics[width=0.49\textwidth]{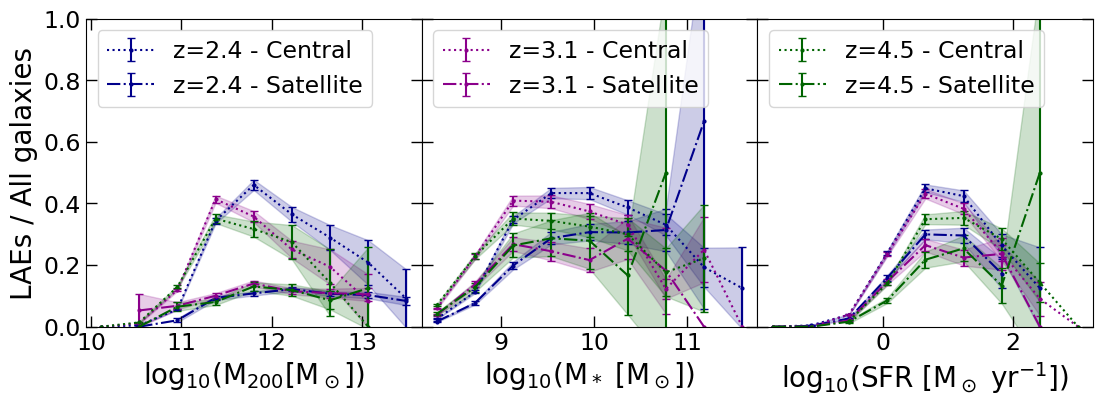}
\end{subfigure}
\caption{
\textit{Upper panel:} Fraction of intrinsic TNG100 LAEs relative to the total galaxy population as a function of different properties, for the three redshifts: $z=2.4$ (blue), $z=3.1$ (magenta), and $z=4.5$ (green). The LAE sample represents the intrinsic LAE population, not affected by our stochastic selection. 
\textit{Left}: Fraction of LAEs as a function of the halo mass, M$_{200}$. \textit{Middle}: dependence with stellar mass, M$_*$. \textit{Right}: as a function of the star formation rate (SFR). Shaded regions correspond to uncertainties, which were estimated based on the dispersion and by assuming Poisson errors. The trends indicate that LAEs preferentially reside in intermediate-mass halos, with moderate stellar masses and SFR, showing a consistent behavior across redshifts.
\textit{Bottom panel:} Same as the upper panel, but showing the fraction of LAEs separated into central galaxies (defined as those with the highest stellar mass within a halo, dotted lines) and satellite galaxies (dashed lines), relative to the total galaxy population in TNG100.
}
\label{fig:frac_lae_gal}
\end{figure}

The upper panel of Figure \ref{fig:frac_lae_gal} illustrates the fraction of LAEs relative to the total galaxies as a function of the logarithm of halo mass (M$_{200}$), stellar mass (M$_*$), and star formation rate (SFR) for the intrinsic TNG100 LAE population. This LAE sample represents the full set of LAEs, unaffected by stochastic detection but obtained by applying the lower luminosity cut for each redshift sample. 
The blue lines represent the population at $z=2.4$, the magenta lines show the population at $z=3.1$, and the green lines indicate $z = 4.5$. The errors were estimated based on error dispersion and by assuming Poisson errors.

The left panel shows that the LAE fraction peaks at intermediate halo masses ($\rm log_{10}$ ($\rm M_{200}/ \rm M_{\odot}) \sim 11.5$) and decreases toward lower and higher masses. As halo mass increases, the relative fraction of LAEs within those halos decreases significantly, by approximately 20$\%$ at the three redshifts. This behavior suggests that Ly$\alpha$ emission is most likely to be observed in galaxies residing in moderately massive halos. Although more massive halos host a larger number of galaxies overall, they may be less efficient at producing LAEs. Therefore, halos with a higher number of galaxies become increasingly difficult to identify using LAEs.
At lower halo masses, the decline in LAE fraction may be linked to inefficient star formation \citep[e.g.,][]{Behroozi2013} or suppression due to photoionization and feedback processes \citep[e.g.,][]{Okamoto2008,Dijkstra2014}. 
In contrast, at higher halo masses, the decrease could be attributed to dust attenuation (indirectly included via the LAE luminosity model) and the presence of more evolved stellar populations that reduce Ly$\alpha$ visibility. 
Notice that these fractions include the LAEs that were removed at random to account for a stochastic chance of the Ly$\alpha$ emission to escape the galaxy, here assumed to be independent of the halo mass.

The middle panel shows a similar trend with stellar mass, where the LAE fraction increases up to $\rm log_{10} (\rm M_*/ \rm M_{\odot}) \sim 10$ ($9.1$ and $9.1$) at $z=2.4$ ($3.1$ and $4.5$), remains roughly constant for a range, and then gradually decreases. Consequently, TNG100 LAEs are more likely to be found among galaxies with intermediate stellar masses.
The drop at high stellar masses reinforces the interpretation that more massive galaxies are likely dustier \cite[e.g.][]{Kumar2025}, which suppresses Ly$\alpha$ visibility. However, the dispersion is higher and the trend is not clear at redshift 4.5.

The SFR shows a similar behavior (right panel): LAEs are more prevalent at intermediate star formation rates, reaching fractions as high as 40$\%$. This trend aligns with theoretical expectations by construction: Ly$\alpha$ emission originates from recombination in HII regions powered by young, massive stars \citep{PartridgePeebles1967,DijkstraWyithe2012}, so the probability of observing a galaxy as a LAE should initially increase with SFR until $\sim10$ $\rm M_{\odot} yr^{-1}$. However, at very high SFRs, the associated increase in dust content and complex gas kinematics is expected to result in enhanced Ly$\alpha$ resonant scattering and absorption \citep[e.g.,][]{Matthee2016}. This is reproduced here by the phenomenological Ly$\alpha$ luminosity model, leading to a reduced LAE fraction. 

Taken together, these trends reveal that LAEs preferentially reside in halos of intermediate mass, as well as intermediate stellar masses and moderate star formation rates. The trends are consistent between redshifts, with a higher LAE fraction observed at $z = 3.1$ and $z=4.5$ compared to $z = 2.4$, particularly at low halo and stellar masses.

In the bottom panel of Figure \ref{fig:frac_lae_gal}, we show the LAE fraction as a function of M$_{200}$, M$_*$ and SFR for the three redshifts, separated into central and satellite galaxies, where a galaxy is defined as the central of its halo if it holds the highest stellar mass. Centrals are depicted with dotted lines, and satellites with dot-dashed lines. We can see that central galaxies exhibit a higher fraction than satellites across most M$_{200}$, M$_*$, and SFR values, at the three redshifts.

The fraction peaks at similar SFRs for both centrals and satellites, suggesting that Ly$\alpha$ emission is related to specific galaxy properties that are less frequent in satellites. This, together with the preference toward lower halo masses shown only by central LAEs, and the similar behavior of the fractions as a function of stellar mass for centrals and satellites, is probably reflecting the influence of the environment on the latter, which is expected to make satellites less able to produce ionizing radiation due to the effect of their environment.

This result reflects the influence of the environment: satellite galaxies are more likely to reside in regions where environmental processes can suppress star formation and remove gas, thereby reducing the likelihood of detectable Ly$\alpha$ emission.

\section{Surface densities of Ly$\alpha$ and UV luminosities normalized by the virial area} 
\label{appen:virial_densities}

\begin{figure*}[!b]
\centering
\includegraphics[width=0.8\textwidth]{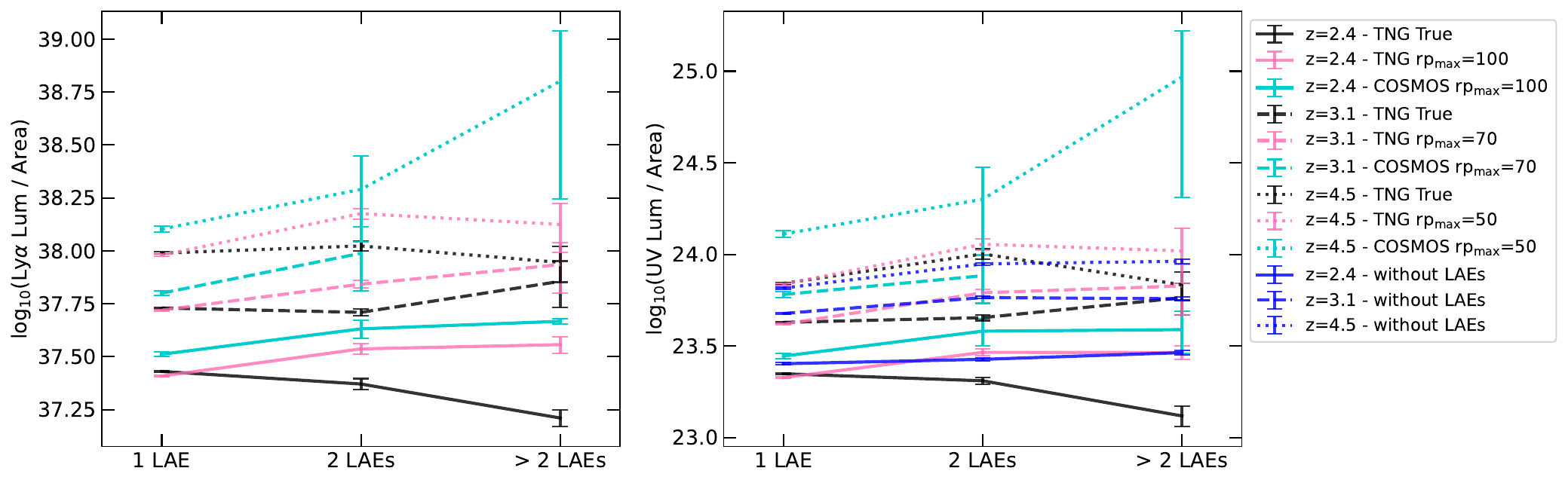}
\caption{Mean virial surface luminosities of Ly$\alpha$ (erg s$^{-1}$ kpc$^{-2}$, \textit{left}) and UV (erg s$^{-1}$ Hz$^{-1}$ kpc$^{-2}$, \textit{right}) as a function of LAE multiplicity. Normalization is based on the area defined by the mean R$_{200}$ of the host halo. Black: TNG100-true; pink: TNG100-identified; cyan: ODIN-COSMOS; blue: halos without LAEs (see explanation in the text). Solid, dashed, and dotted lines correspond to $z = 2.4$, $z = 3.1$, and $z = 4.5$, respectively. Error bars show the standard error of the mean in each bin.}
\label{fig:virial_densities}
\end{figure*}

As shown in the right panels of Figure \ref{fig:lum_uv_a_ngal}, Figure \ref{fig:virial_densities} shows the the mean virial surface luminosities of Ly$\alpha$ (erg s$^{-1}$ kpc$^{-2}$, left panel) and UV (erg s$^{-1}$ Hz$^{-1}$ kpc$^{-2}$, right panels) as a function of LAE multiplicity. In this case, the normalization is based on the area defined by the mean R$_{200}$ (radius of the host halo whose mean density is 200 times the critical density of the universe) obtained from the mock, applied to both mock and observations. The black lines correspond to the TNG100-true values, the pink lines to the TNG100-identified sample, and the cyan lines to the ODIN-COSMOS data, with solid, dashed, and dotted lines representing $z = 2.4$, $z = 3.1$, and $z = 4.5$, respectively. Error bars indicate the standard error of the mean in each bin. In most cases, higher LAE multiplicity is associated with enhanced surface brightness densities.

In the right panel, we also show the expected densities of halos without LAEs in blue for the three redshifts. These halos have radii within the ranges defined by the mean radii of the identified isolated systems, the pairs of LAEs, and the triplets of LAEs, including the corresponding data dispersion, respectively. We find that the same trend holds even for halos without Ly$\alpha$ emitters, indicating that our LAE-selected halos trace the UV density in an unbiased way.

\end{appendix}

\end{document}